\documentclass[12pt,a4paper,dvips]{article}
\usepackage{a4p}
\usepackage{cite,mcite}
\usepackage{graphicx,rotating}
\usepackage{physics }
\usepackage{epsfig }
\usepackage{l3_titlePH,ifthen}
\usepackage{Lep}
\date{March 3, 2004}
\preprint{2004-005}

\usepackage{Lep}
\Lep{2}

\def\Journal#1#2#3#4{{#1} {\bf #2} (#4) #3}

\def\NIMA{{Nucl. Instr. Meth.} A}

\def\PLB{{Phys. Lett.}  B}
\def\PRL{Phys. Rev. Lett.}
\def\PRD{{Phys. Rev.} D}
\def\ZPC{{Z. Phys.} C}
\def\PRP{{Phys. Rep.} C}
\def\CPC{Comput. Phys. Commun.}

\newlength{\capwidth}
\def\pb{\mbox{pb$^{-1}$}}
\newcommand{\EE}{\mathrm{e}^+\mathrm{e}^-}

\newcommand{\gamgam}{\gamma \gamma^*}

\newcommand{\roc}{\mathrm{\rho^\pm}}
\newcommand{\rocroc}{\mathrm{\rho^+\rho^-}}
\newcommand{\ro}{\mathrm{\rho^0}}
\newcommand{\roro}{\mathrm{\rho^0\rho^0}}
\newcommand{\pipi}{\mathrm{\pi^+\pi^-}}
\newcommand{\piz}{\mathrm{\pi^0}}
\newcommand{\pizpiz}{\mathrm{\pi^0\pi^0}}
\newcommand{\q}{ Q^2 }
\newcommand{\mgg}{ W_{\gamma \gamma }}
\newcommand{\ptt}{ p_t^2 }

\begin{document}

\begin{titlepage}
\title{Measurement of Exclusive  $\boldmath{\rho^+\rho^-}$ Production \\
in High-$\mathbf{Q^2}$ Two-Photon Collisions   at LEP}

\author{The L3 Collaboration}


\begin{abstract}

Exclusive $\rocroc$ production in two-photon 
collisions  involving a single highly-virtual photon  
is studied for the first time with data collected by the L3 experiment at LEP 
  at centre-of-mass energies
\mbox{$89 \GeV < \sqrt{s} < 209 \GeV{}$} with a total integrated 
luminosity of $854.7~\pb$. 
The cross section  of the process  $\, \gamma \gamma^* \rightarrow \rocroc \,$  
is determined as a function of the photon virtuality, $\q$,
and the 
two-photon centre-of-mass energy, $\mgg$,
in the kinematic region:
$1.2 \GeV^2 < \q < 30 \GeV^2$ and 
$1.1 \GeV < W_{\gamma\gamma} < 3 \GeV$. 
The  $\rocroc$ production cross section is found to be of the same magnitude
as the cross section of the process $\gamma \gamma^* \rightarrow\roro$,
measured in the same kinematic region by L3, 
and to have similar $\mgg$ and $\q$ dependences.

\end{abstract}

\submitted

\vfill

\end{titlepage}


\section {Introduction}
In this Letter, we present the first measurement of the process:
\begin{eqnarray}
\label{eq:eqn01}
\EE \to \EE \gamgam \to \EE \rocroc
\end{eqnarray}
\noindent
in a kinematic region of large momentum transfer, obtained 
with data collected by the L3 detector\cite{L3} at LEP.
 In this  kinematic domain,  one of the interacting photons, $\gamma$,  is quasi-real
 and the other, $\gamma ^*$, has a large virtuality, $\q$,  
defined by a scattered electron\footnote{Throughout this Letter,
the term ``electron'' denotes both electrons and positrons.}
detected (``tagged'') in the forward electromagnetic calorimeter, 
used to measure the luminosity.
This work continues our study  of exclusive $\gamgam \ra \rho \rho$  production:
 our measurement of the process $\gamma\gamma^*  \to \roro$ was
recently published\cite{L3paper269} and here the charge-conjugate channel is analysed. 
The $\gamma\gamma \ra \rho ^+ \rho ^- $ exclusive production was 
previously studied only at low $\q$ for  quasi-real photons\cite{ARGUS,CELLO}.
 \par
 The interest in  exclusive production of hadron pairs in two-photon 
interactions at high momentum transfer is due to recently developed 
methods for calculating the cross section of such processes
in the framework of  perturbative QCD\cite{QCD}. In these models, the
exclusive process is factorised into a perturbative, calculable,
short-distance scattering $\rm \gamma\gamma^*\rightarrow q\bar{q}$ or 
 $\rm \gamma\gamma^*\rightarrow gg$ and non-perturbative matrix elements
describing the transition of the two partons into hadron pairs, which are
called generalised distribution amplitudes.
A comprehensive theoretical analysis of our $\gamma\gamma\rightarrow\roro$ data\cite{L3paper269} in this framework 
was recently performed\cite{ANIKIN}.

\par  
 The squared four-momentum transfer, $\q$,
is determined by the beam energy, $ E_b $, and
the energy and scattering angle of the tagged electron,
$ E_s $ and $ \theta_s $, by the relation:
 \begin{eqnarray}
 \label{eq:eqn02}
 \q = 2 E_b E_s (1 - \cos \theta_s)  .
\end{eqnarray}


The bremsstrahlung production  of $\rocroc$ pairs, which represents a background to the
process (\ref{eq:eqn01}), is strongly suppressed in the
kinematic region of our measurement \cite{DIEHLPAP,DIEHL}.

%
%
%

The data used in this study, the kinematic regions covered  and the analysis 
techniques employed are similar to those of  our measurement of $\roro$ production
in tagged two-photon interactions \cite{L3paper269}.
The data  correspond  to an integrated luminosity of 854.7 \pb , 
out of which   
148.7  \pb~were collected at $\EE$ centre-of-mass energies, \rts , around the Z resonance (Z pole),
 and 706.0 \pb {} at 
$161 \GeV \leq \rts < 209 \GeV$ (high energy), corresponding to an average \rts {} of 195 \GeV.
 The production cross section is determined as a function
of the invariant mass of the hadronic system, $\mgg$, and 
as a function of $\q$ in the kinematic region defined by the intervals:
\begin{equation}
\label{eq:rangelep1}
1.2 \GeV^2  <  \q  <8.5 \GeV^2 \,\,\,\, \mathrm{(Z \,\,pole)};
\end{equation}
\begin{equation}
\label{eq:rangelep2}
8.8 \GeV^2  <  \q  <30 \GeV^2 \,\,\,\,\,   \mathrm{(high \,\, energy)};
\end{equation}
\begin{equation}
\label{eq:rangewgg}
1.1 \GeV  <   \mgg   <3 \GeV .\,\,\,\,\, 
\end{equation}

The results are compared to our measurement of $\roro$ production at high $\q$ and to the
Vector Dominance model\cite{GINZBURG}, as well as to the expectations of a QCD model\cite{DIEHLPAP}.


\section {Experimental considerations}
The L3 detector is described in detail in Reference~\citen{l3big}. The
sub-detectors used for the study of the reaction (\ref{eq:eqn01})  
are the charged-particle
tracker, the electromagnetic calorimeter and the small-angle
luminosity monitor.  For this analysis,
their fiducial volumes and thresholds are chosen so as to
achieve the necessary resolution and  background rejection, as discussed in the following.
\par 
The central detector is a cylindrical high resolution drift chamber, complemented by a
silicon micro-vertex detector near  the beam pipe, in a magnetic 
field of 0.5 T.  A polar-angle fiducial volume is chosen as $15^\circ \le \theta \le 165^\circ$.
The transverse momentum resolution is parametrised as  
$ \sigma _{p_t} / p_t = 0.018 \; p_t ($\GeV$) \oplus 0.02$.
Only tracks which 
 come from the interaction vertex,  have
transverse momentum greater than  $100 \MeV $ and  an energy loss in the tracking chamber
compatible with the pion hypothesis are considered in this analysis. 
\par
The electromagnetic calorimeter consists of an array of 10734 BGO crystals, with the form 
of a truncated pyramid of $2 \times 2 \,\, \mathrm{cm ^2}$ base. The crystals are arranged in two half barrels
with a polar angle coverage
$ 42^\circ \le \theta \le 138^\circ $ and in two end-caps covering $11.6^\circ \le \theta \le 38^\circ$
and $142^\circ \le \theta \le 168.4^\circ$. 
The material preceding the barrel part of the electromagnetic calorimeter, amounts to 20\% 
of a radiation length, increasing to 60\% of a radiation length in the endcap regions.
The energy resolution, $\sigma_{\mathrm E} /{\mathrm E}$, varies from 5\% at 50 \MeV {} 
to about 1\% for energies greater than 10 \GeV. In the following, only showers with energy greater than 60 \MeV {} are considered for $\pi^0$ reconstruction.
\par
The luminosity monitor, installed on each side of the detector and also made out of  BGO crystals, 
covers  the polar angle range
 $25 \;\mathrm {mrad} \le \theta \le 68 \;\mathrm {mrad}$ for the Z-pole  runs and 
  $31 \;\mathrm {mrad} \le \theta \le 65 \;\mathrm {mrad}$ for the high-energy runs,
   when a mask was introduced to protect the detector from the beam halo.



\section {Event selection}

The reaction (\ref{eq:eqn01}), contributing to the process
\begin{eqnarray}
\label{eq:eqn06}
\EE \to  \mathrm{e^+ e^-_{tag}}\pi^+\pi^-\pi^0\pi^0 ,
\end{eqnarray}
is identified  by a scattered beam electron, $ \mathrm{e_{tag}}$,
detected in the luminosity monitor, two charged pions measured in the tracking chamber, 
and energy clusters from the two-photon decays of the $\pi^0 $'s
deposited in the electromagnetic calorimeter.
These events  are accepted by several independent triggers with  major
contributions coming from a charged-particle trigger \cite{L3T},
with different features for the  Z-pole and high-energy data-taking periods, 
and an energy trigger demanding a large energy deposition in the luminosity
monitor in coincidence with at least one track \cite{SEtrig}.
The combined trigger efficiency, as determined from the data itself, is
$(85.2 \pm 3.8)\% $ at the Z pole and 
$(96.8 \pm 1.5)\% $ at   high energy.

Single-tagged events are selected by requiring an electromagnetic cluster with energy 
greater then  $80\% $ of the beam energy   reconstructed in the luminosity monitor.

\begin{figure} [p]
  \begin{center}
   \vskip -1.0cm
     \mbox{\epsfig{file=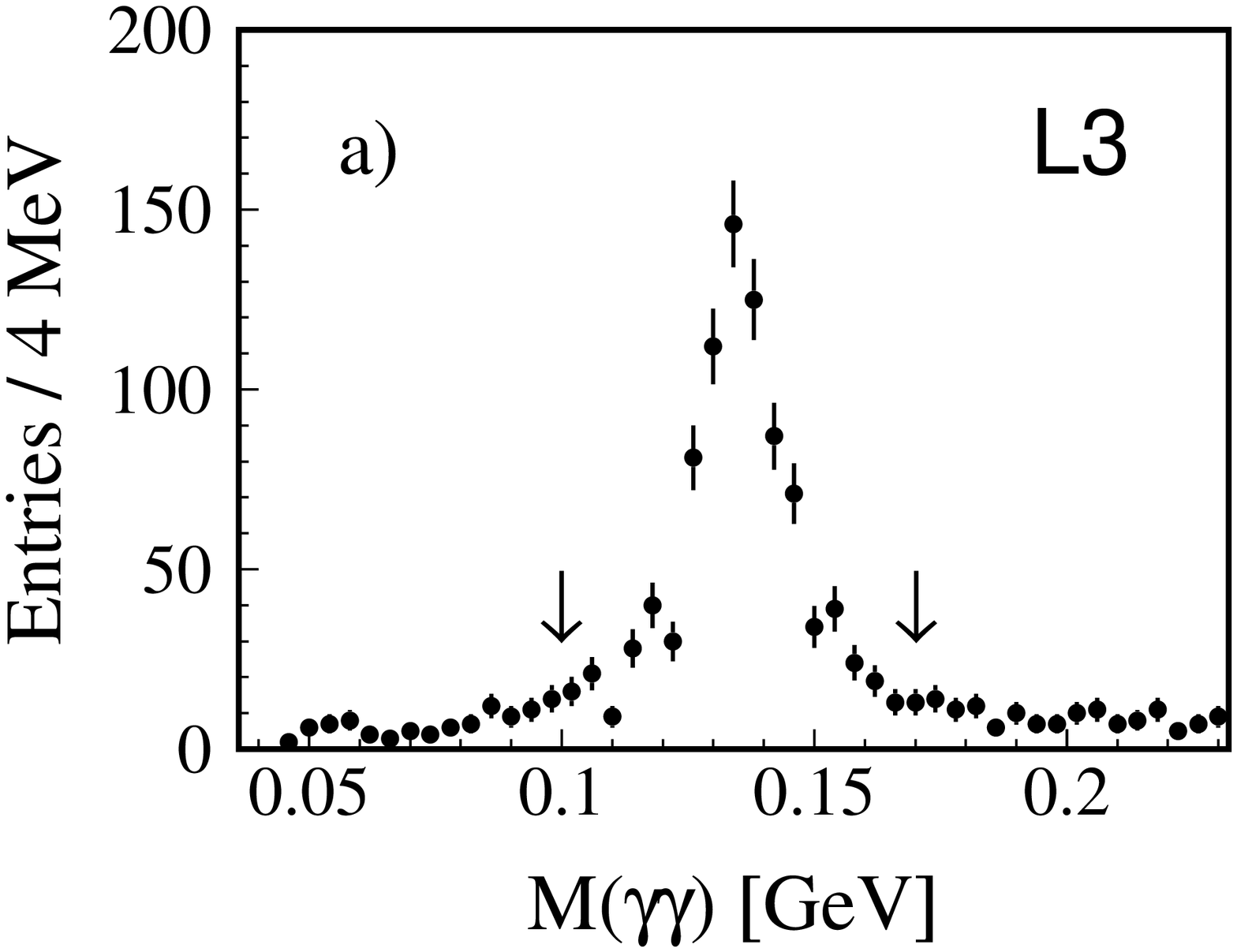,width=0.51\textwidth}}
     \mbox{\epsfig{file=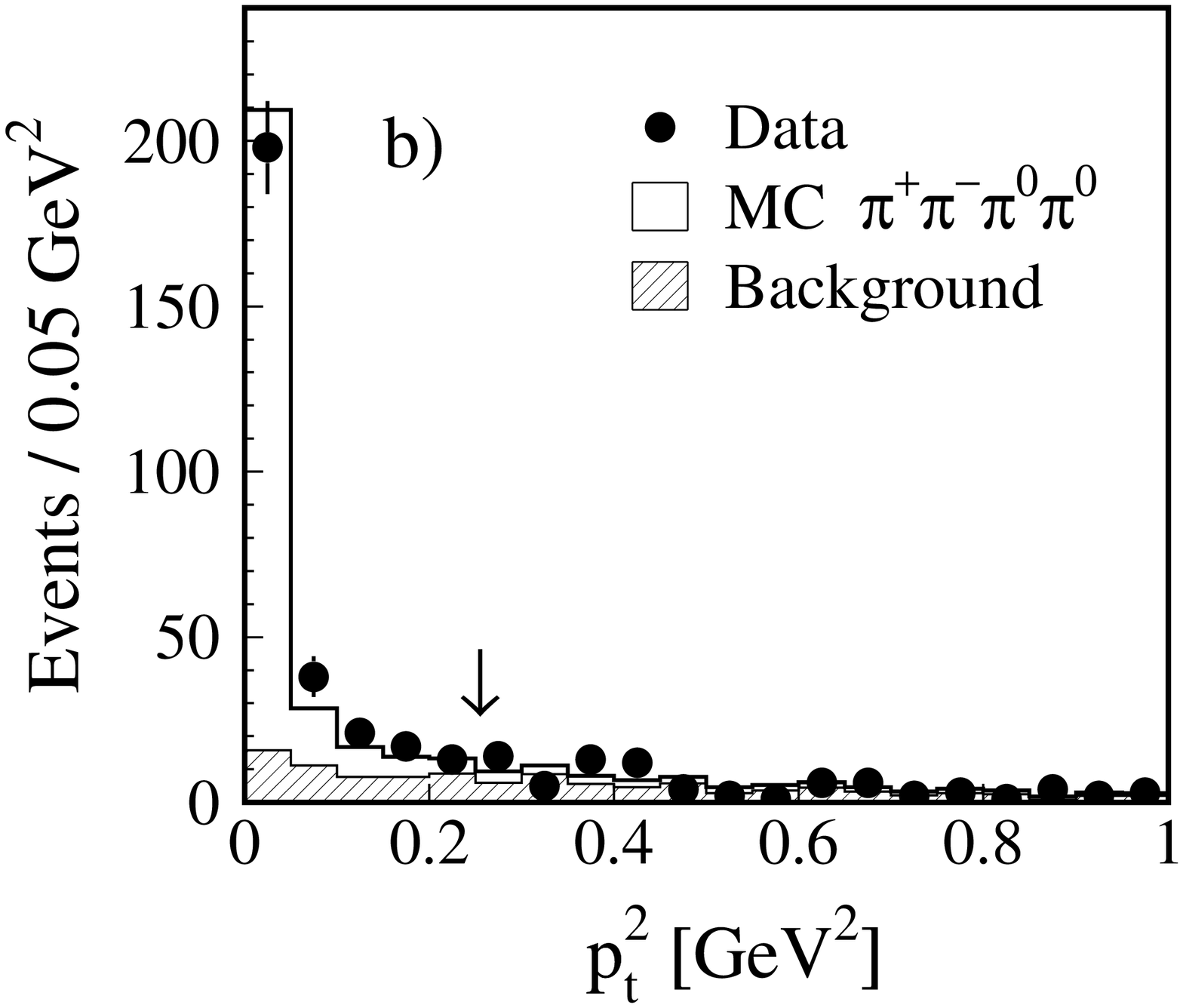,width=0.51\linewidth}}
     \mbox{\epsfig{file=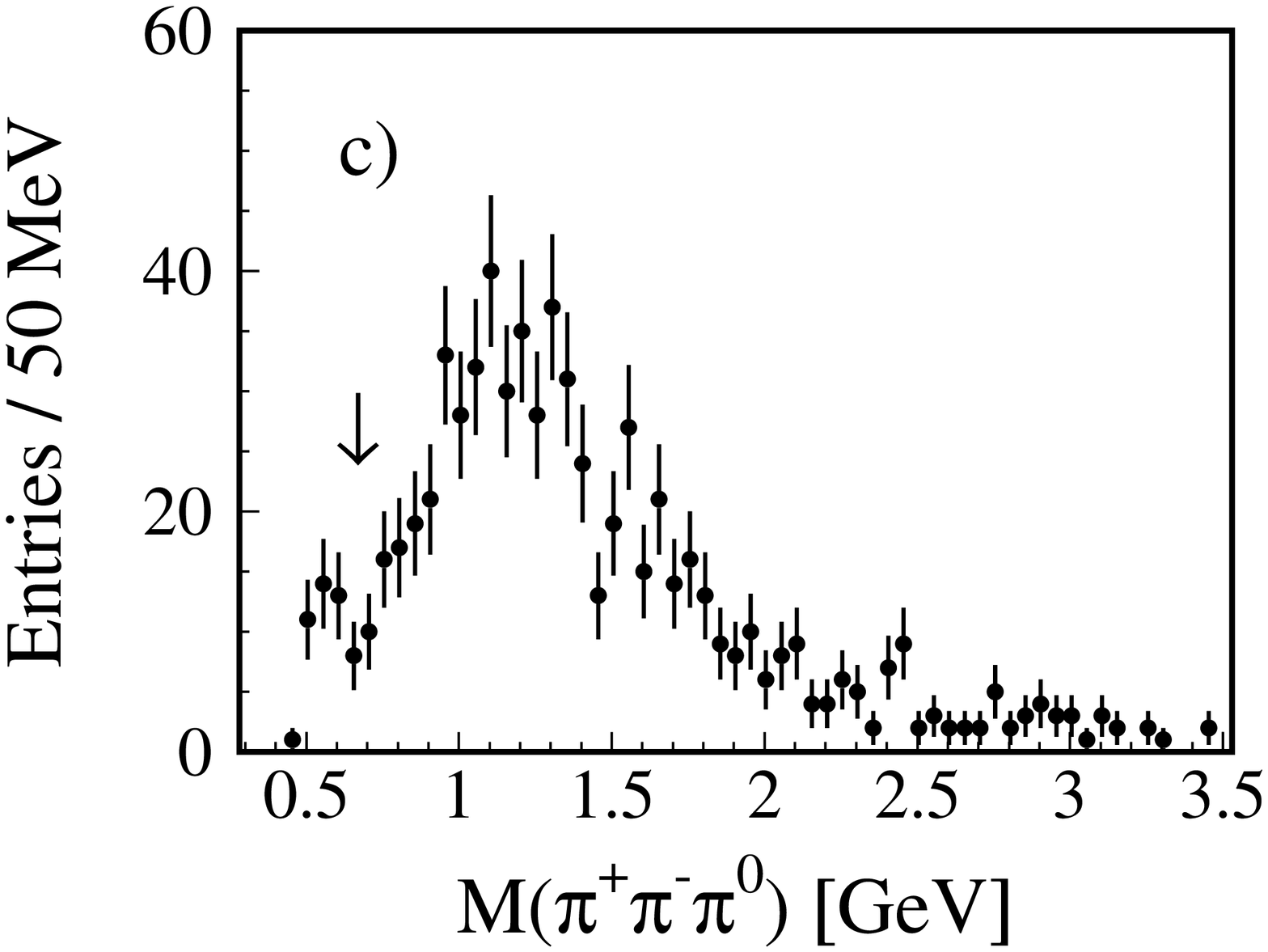,width=0.51\textwidth}}
   \vskip -0.5cm
  \end{center}
  \caption[]{ Observed distribution of 
             a) the two-photon effective mass  
             (two entries per event); b) the event $\ptt$ for $1.1 \GeV < \mgg < 3 \GeV$;
             c) the mass of the $\pipi\piz$ system (two entries per event).
               Monte Carlo simulation of four-pion events (open histogram) and the background
	        estimated from the data (hatched histogram) are also shown in b).
             The arrows indicate the selection cuts.
            }
\label{fig:fig1}
\end{figure}

The event candidates must have exactly two tracks with zero total charge 
 and four or five photons, 
identified as isolated clusters in the electromagnetic calorimeter,
not matched with a charged track.
Photons  are paired to  reconstruct
 neutral pions and their effective mass must be 
between $100 \MeV$ and $170 \MeV$, as shown in Figure~\ref{fig:fig1}a.
To improve the resolution of the reconstructed  $\piz$ four-momentum,
a constrained 1C kinematic fit to the nominal $\piz$ mass is performed for
each $\piz$ candidate. If more than one $\pizpiz$ combination exists in an event, 
the one with smallest sum of the $\chi^2$ from the  constrained fits of its 
constituent $\piz$'s is taken. 
Events which contain an additional photon candidate, not used in the
selected $\pizpiz$ pair, are retained only if the energy of that
photon is less than 300 \MeV{} and does not exceed $10\%$ of the
energy of the selected $\pizpiz$ combination. Allowing for these
additional soft photon increases the acceptance. These ``noise''
photons are due to instrumental sources, to beam-related backgrounds or
remnants of hadronic showers.

To ensure that an exclusive final state is detected, 
the  momenta of the tagged electron and the four-pion system must
be well balanced in the plane transverse to the beam direction. 
The total transverse momentum squared, $\ptt $, of the four-pion final state 
and the scattered electron, shown in Figure~\ref{fig:fig1}b, is required to be less than $ 0.25 \GeV^2$.

 Figure~\ref{fig:fig1}c shows the mass spectrum of the $\pipi\piz$ subsystem of the four-pion 
final state.  Apart from an $\eta$ signal near the kinematic threshold, no other resonance 
structure is visible. 
Final states containing $\eta$'s represent a background to the process (\ref{eq:eqn01}) 
and are  removed  by requiring the three-pion mass to be above 0.65 \GeV.

\begin{figure}[p]
\begin{center}
   \vskip -1.5cm
{\epsfig{file=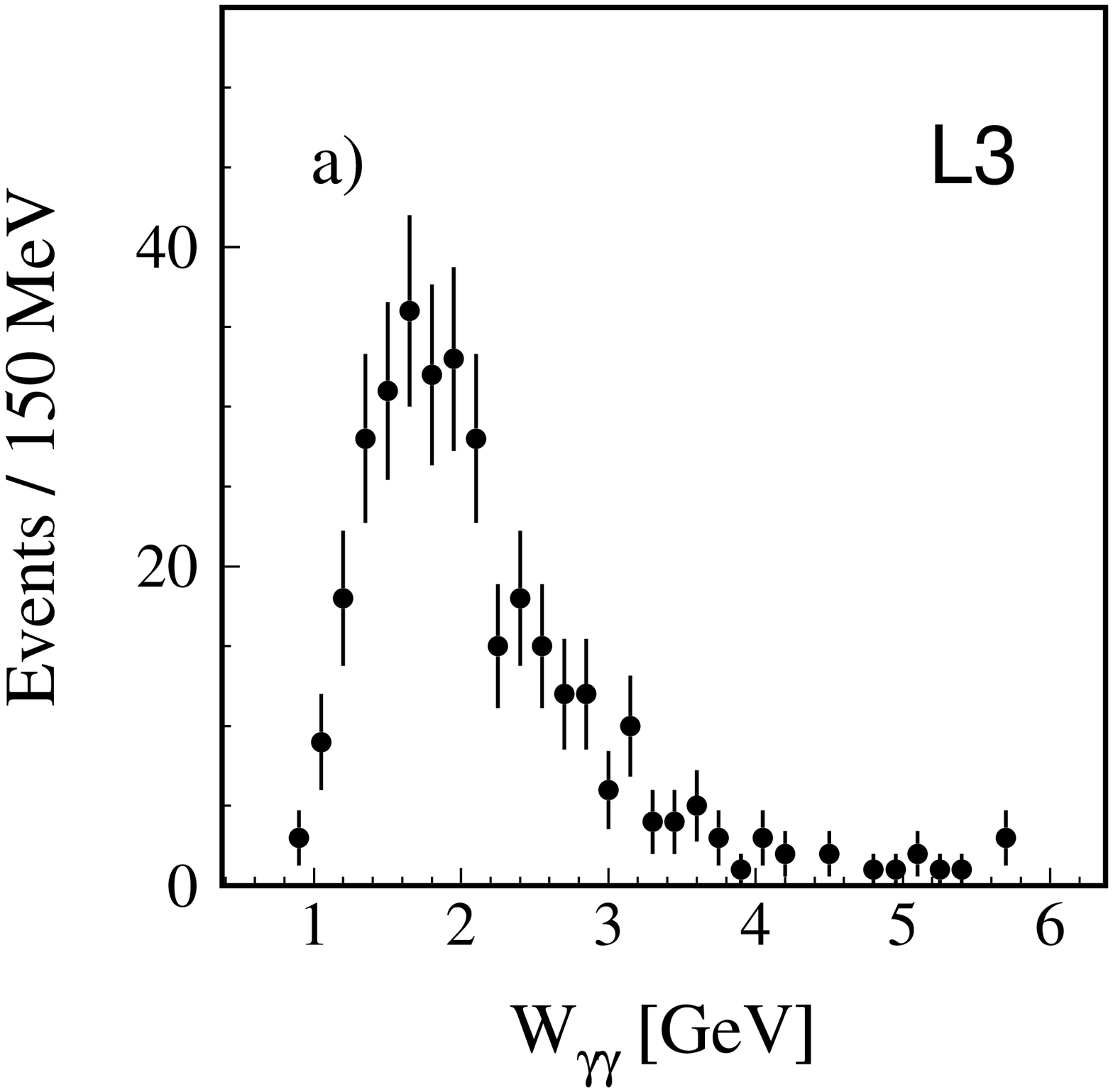,width=0.49\linewidth}}
{\epsfig{file=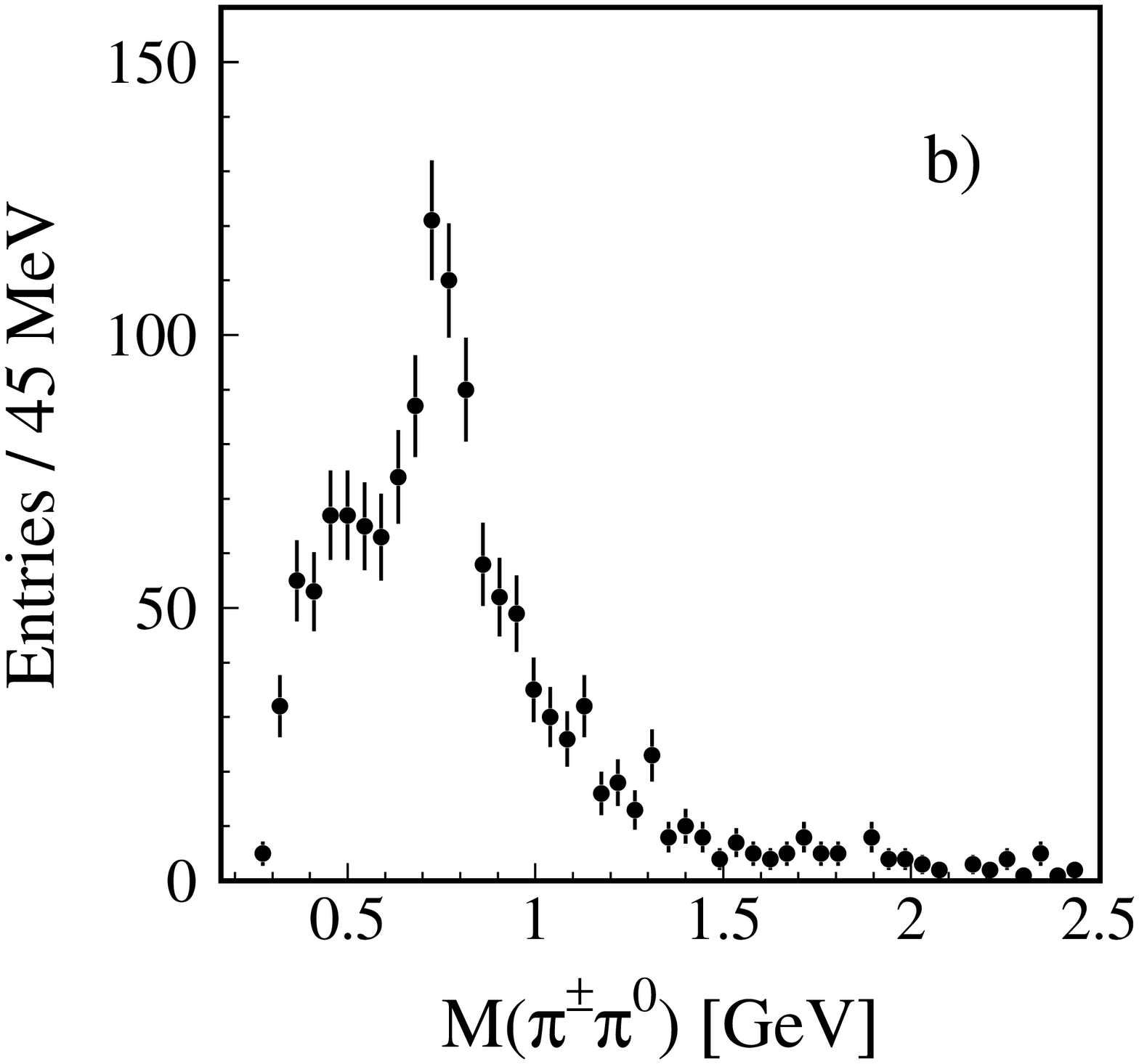,width=0.49\linewidth}}
{\epsfig{file=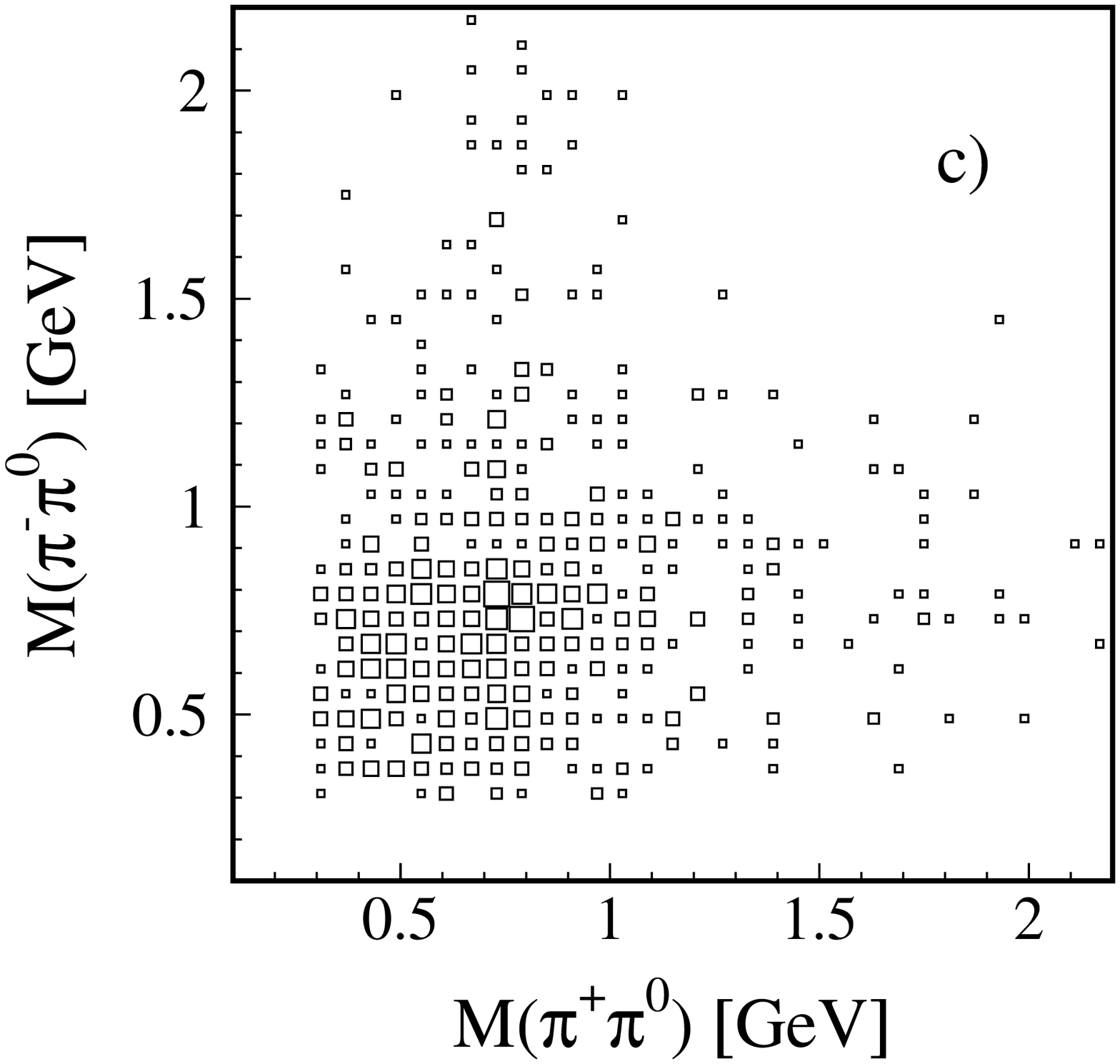,width=0.49\linewidth}}
{\epsfig{file=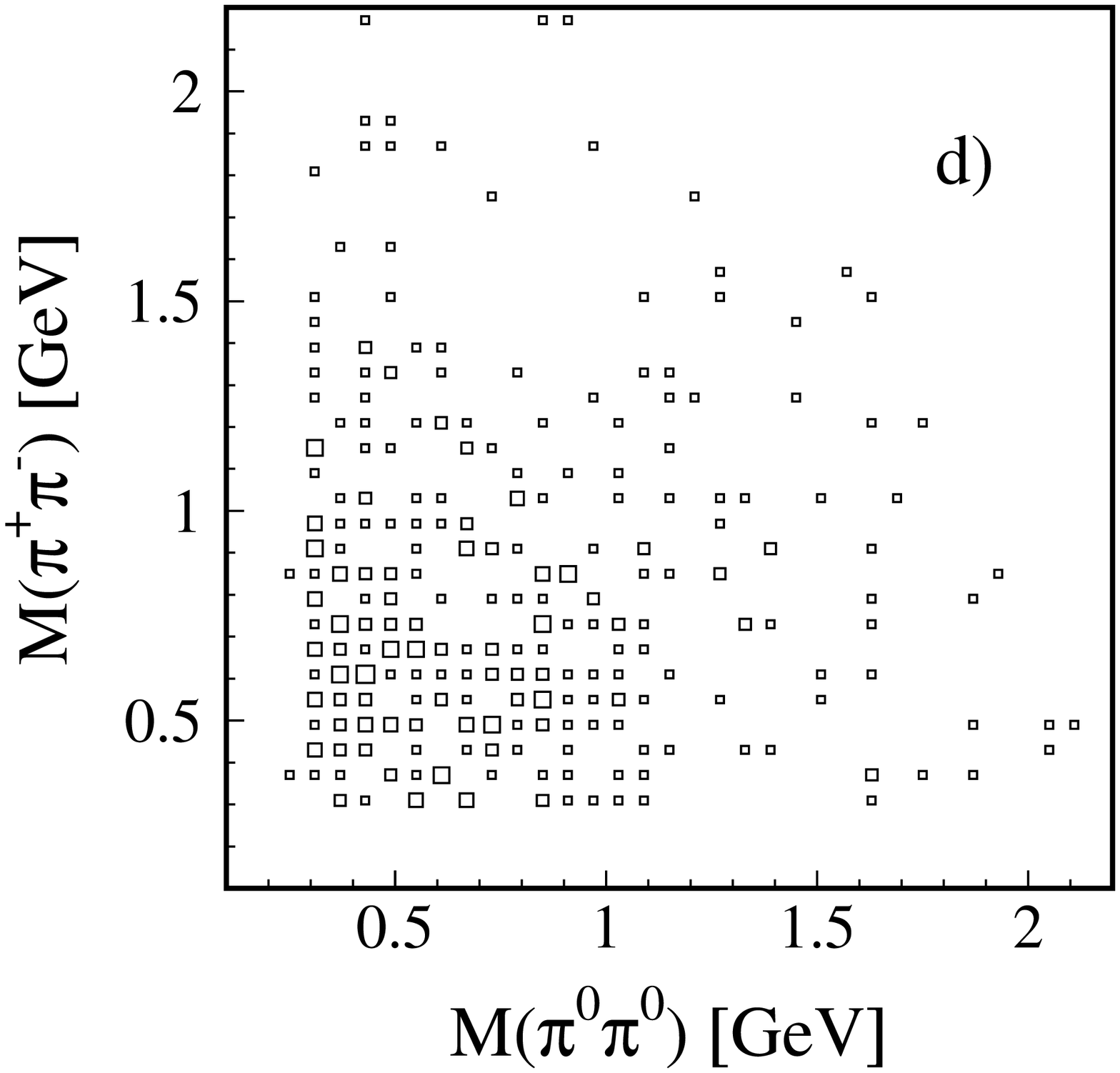,width=0.49\linewidth}}
\end{center}

  \caption[]{Effective mass distributions for the selected events:
             a) mass of the four-pion system, $\mgg$,
             b) mass of $\pi^\pm \piz$ combinations (four entries per event),
             c) correlation between the masses of the $\pi^-\piz$ and  $\pi^+\piz$ 
                 pairs (two entries per event),
             d) correlation between the masses of the $\pipi$ and  $\pizpiz$ pairs.
             The two-dimensional distributions have a bin width of 
             $60 \times 60 \MeV^2$, the size of the boxes is proportional to the number 
             of entries and both plots have the same vertical scale.
           }
\label{fig:fig2}
\end{figure}

After all cuts,  343 events are observed, out of which 
224 events are at the  Z pole  and 119 events are at high energy.  
The four-pion mass spectrum of these
events is shown in Figure~\ref{fig:fig2}a. The mass distribution of the $\pi^\pm\piz$
combinations of the selected events, shown in Figure~\ref{fig:fig2}b,
shows a peak at the   $\rho$ mass.
A clustering of entries is observed at the  crossing of the $\rho^\pm$ 
mass bands in the correlation plot of the masses of the charged $\pi^\pm\piz$ combinations,
shown in Figure~\ref{fig:fig2}c. No resonance structure is observed in Figure~\ref{fig:fig2}d
for  the correlation plot of the masses of the $\pipi$ and $\pizpiz$ combinations. 
These features of the two-particle mass correlations give evidence for a signal from
$\rocroc$ intermediate states.

\section {Background estimation}

The  contribution to the selected sample due to $\EE$   annihilation   is  negligible.
The  background  from tagged exclusive $\pipi\piz$ final states,  where  
  photon candidates due to noise mimic the second $\piz$,
is also negligible, as found  by studying the $\ptt$ distribution of  
$\mathrm {e_{tag}}\pipi\piz$ subsystems of the selected events.

Two sources of background remain: 
partially reconstructed events with higher particle multiplicities where tracks 
or photons escape  detection  
and signal events 
with one or more photons  substituted by  photon candidates due to noise.
The latter  has a component of its 
$\ptt$ distribution  similar to that   of the signal. These backgrounds are studied directly with the data.

To estimate the background due to feed-down from higher-multiplicity final states, 
we select data samples of the type $\pi^\pm \pi^\pm \pizpiz$. In addition, we select 
  $\pipi\pizpiz\piz$ events and exclude one $\piz$ from the reconstruction. 

An event-mixing technique is employed in order to reproduce events from the second
background source: 
 one or two photons forming
a $\piz$   are excluded  from a selected event and replaced by photons from 
another data event. 

All these events are required to pass the event selection procedure discussed above, 
with the exception of the charge-conservation requirement
for the $\pi^\pm \pi^\pm \pizpiz$ sample. 
For the $\pipi\pizpiz\piz$ events only the $\pipi\pizpiz$ subsystem is considered.
The $\ptt$ distributions of the accepted background-like data events
are combined with the distribution of selected  $\pipi\pizpiz$  Monte Carlo events
so as to reproduce the measured $\ptt$ distribution of the selected data events, as shown in 
Figure~\ref{fig:fig1}b.
The contribution of the background from partially reconstructed events is 
on average three times higher than the second background.
The result of this procedure, applied for the events in the kinematic region defined by
the conditions (\ref{eq:rangelep1}), (\ref{eq:rangelep2}) and (\ref{eq:rangewgg}), 
is shown in Figure~\ref{fig:fig1}b and the background levels
 are quoted in Tables~\ref{tbl:xsectq2}--\ref{tbl:xsectwgg_lep2}.
 \par
 To estimate the uncertainties on the background correction, the background
 evaluation procedure is repeated by excluding, in turn, each of the
background-like data samples. The larger value between the statistical uncertainty on the background determination 
and the observed variation in the background levels is retained as
uncertainty. It  varies in the
range \mbox{$4\% - 8\%$}.


\section {Data analysis}

The $\rocroc$ production  is studied in bins of  $\q$ and $\mgg$. 
These variables are reconstructed with a resolution better than  3\% and the 
 chosen bin widths  are such  that 
 the event migration between adjacent bins is negligible. 
The production cross section is determined in the restricted  $\mgg$-region (\ref{eq:rangewgg}),
which contains 287 events, of which 195 events are at  the Z pole  and 92 events are at  high energy.


\subsection {Production model}

To estimate the number of $\rocroc$ events in the selected four-pion data sample,
we consider non-interfering contributions from three processes: 
\begin{eqnarray}
\label{eq:eqn04}
&&  \gamgam \to \rocroc;    \nonumber \\
&&  \gamgam \to  \roc\pi^\mp \pi^0;  \\
&&  \gamgam \to \pipi \pizpiz ,\, \rm{non-resonant.} \nonumber  
\end{eqnarray}
\par
Our data do not show any evidence for sub-processes involving
production of high-mass resonances. However, 
the $\roc\pi^\mp \pi^0$ term can absorb possible
contributions from intermediate states containing $a_1(1260)$ and $a_2(1320)$ 
resonances.

Monte Carlo samples of the processes (\ref{eq:eqn04}) are generated
with the  EGPC~\cite{LINDE} program. About 6 million events of each sub-process
are produced for both the  Z-pole and the high-energy regions.
The $\mgg$ and $\q$ dependences  are those of the  
$\gamma\gamma$ luminosity function~\cite{BUDNEV} 
and  only isotropic production and phase-space decays are included.
These events are  processed in the same way as the data, 
introducing specific detector inefficiencies  for the different data taking periods. 
For acceptance calculations, the Monte Carlo events are assigned a $\q$-dependent weight,
evaluated using the  GVDM~\cite{GVDM} form-factor for both photons.
The detection efficiencies, calculated taking into account the detector acceptance 
and the efficiency of the selection procedure, are 
listed in Tables~\ref{tbl:xsectq2}--\ref{tbl:xsectwgg_lep2}.
They are in the range of $3\% - 6\%$ ,  very similar for all sub-processes. The efficiency is 
mostly limited by the kinematics of the two-photon reaction which boosts
the hadronic system along the beam direction.
The geometrical coverage of the electromagnetic-calorimeter
fiducial-volume affects the photon acceptance and thus  the efficiency
for $\pi ^0$ reconstruction.

\begin{figure}[p]
\begin{center}
{\epsfig{file=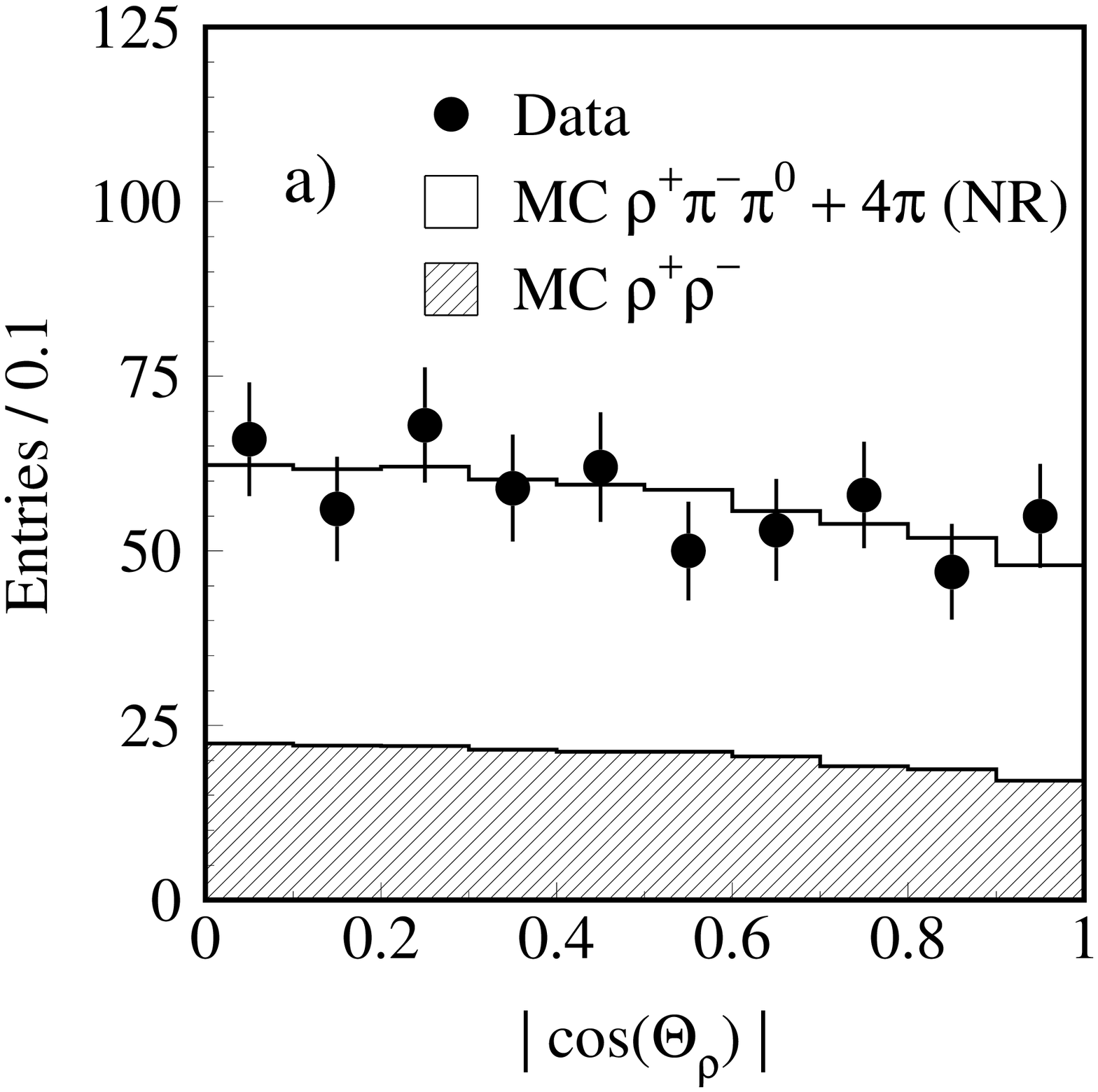,width=0.49\linewidth}}
{\epsfig{file=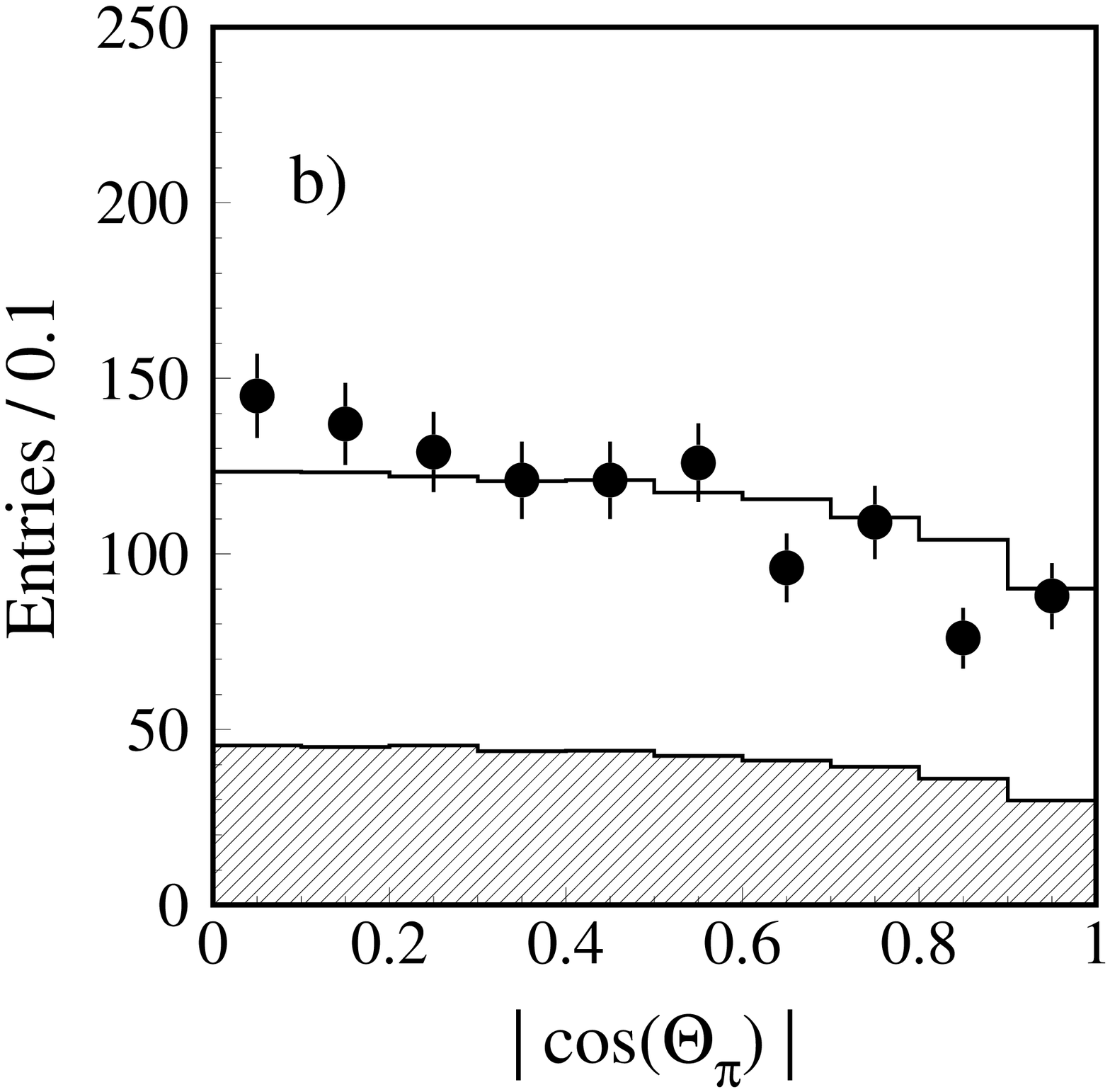,width=0.49\linewidth}}
{\epsfig{file=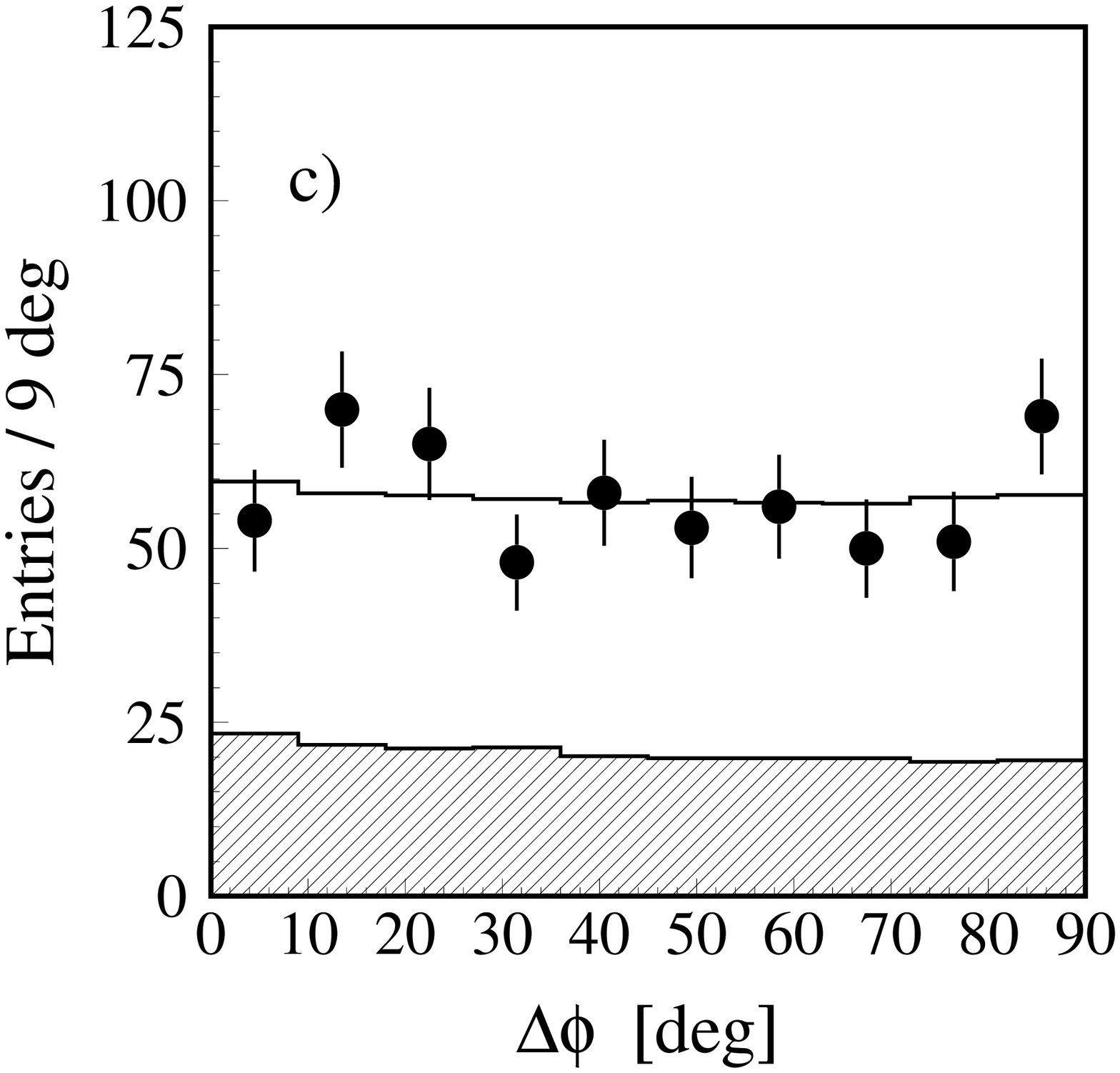,width=0.49\linewidth}}
{\epsfig{file=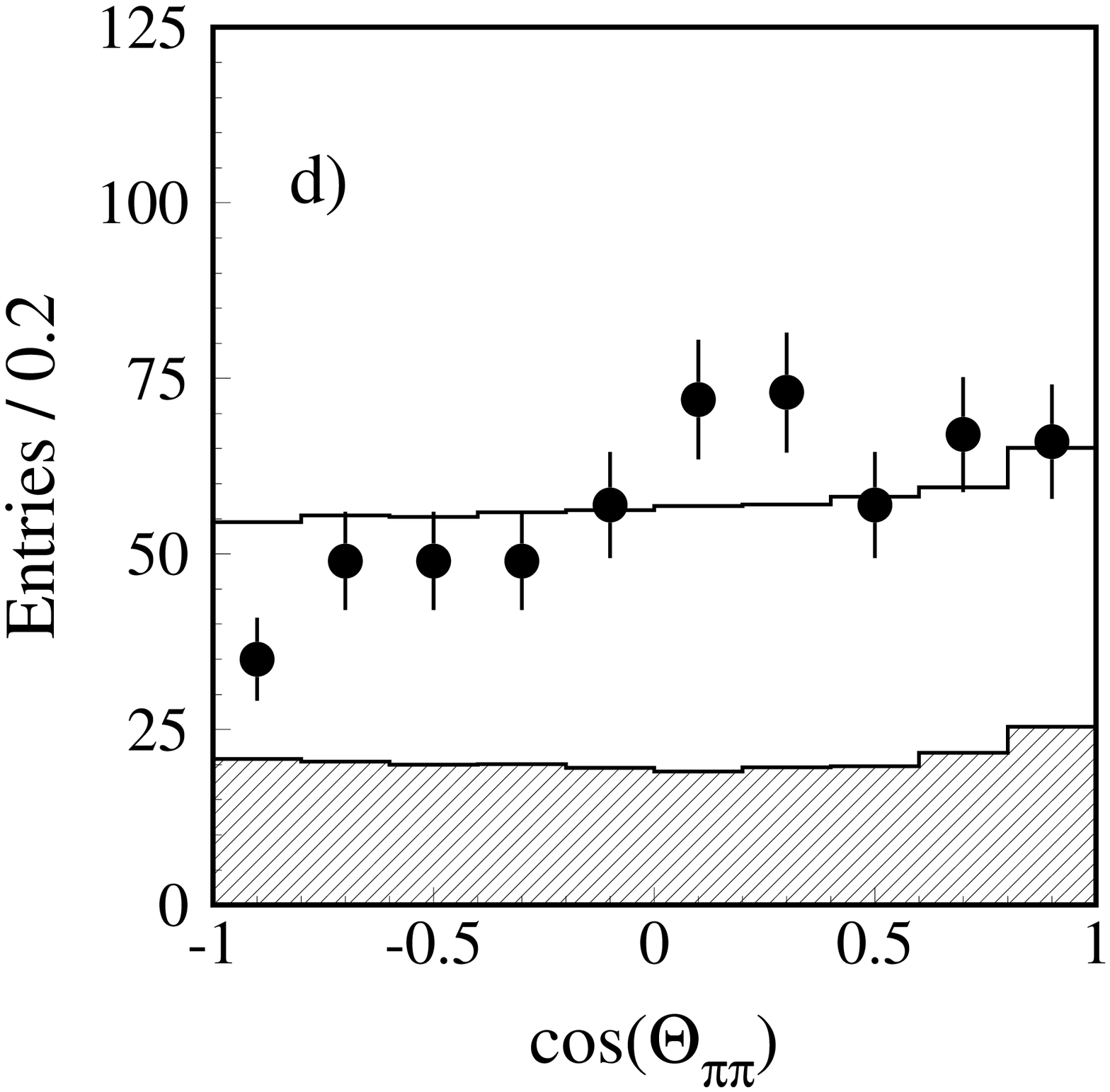,width=0.49\linewidth}}
\end{center}
  \caption[]{Comparison of data   and   Monte Carlo   angular distributions:
           (a) $\mid \cos \Theta_\rho \mid$,  
           the cosine of the polar angle of
           the $\roc$ with respect to the two-photon axis in the two-photon
           centre-of-mass system;
           (b) $\mid \cos \Theta_\pi \mid$, 
           the cosine of the polar angle of the 
           charged pion in its parent $\roc$ helicity-system;
           (c) $\Delta \phi$, the angle between the decay planes of
           the $\rho^+$ and $\rho^-$ 
           mesons in the two-photon centre-of-mass system;
           (d) $ \cos \Theta_{\pi\pi} $, 
           the cosine of the opening angle between
           the  $\pi^+$ and $\pi^-$ directions of flight, each one defined in its parent $\roc$
           rest-system.
           There are two entries per event in (a), (c) and (d) and four entries per event in (b).
           The points represent the data, the hatched area shows  the $\rocroc$ component
    and the open area shows the sum of
             $\roc\pi^\mp \pi^0$ and $\pipi\pi^0\pi^0$ (non-resonant) components.
           The fraction of the different components are determined by the fit and the
   total normalisation is to the number of the events.

           }
\label{fig:angles}
\end{figure}

The efficiency is found to be uniform in the two-photon centre-of-mass
system and it is therefore insensitive to the details of the Monte
Carlo production model. The angular distributions of the reconstructed Monte Carlo events
are similar to the generator-level ones and in good agreement with
those observed in data, as shown in Figure~\ref{fig:angles}.



\subsection {Fit method}

The set, $\Omega$, comprising the six two-pion masses in an event, namely
the four charged  combinations  $\pi^\pm\piz$ and the two neutral combinations, $\pipi$ and $\pizpiz$,
provides a complete description of a four-pion event in our model of isotropic
production and decay.
For each data event, $i$, with measured variables $\Omega_i$, we calculate the
probabilities, 
$ P_j(\Omega_i)$, that the event resulted
from   the production mechanism  $j$.
A likelihood function is defined as:
\begin{eqnarray}
\label{eq:eqn05}
&& \Lambda = \prod_{i}   \sum_{j=1}^{3} \lambda_j P_j (\Omega_i) 
 \,,\;\;\;\;\;\; \;\;\sum_{j=1}^{3} \lambda_j  = 1,
\end{eqnarray}
where $\lambda_j$ is the fraction of the process $j$ in the
$\pipi\pizpiz$ sample for a given  $\q$ or $\mgg$  bin
and  the product runs over all data events in that bin.
The probabilities $P_j$ are determined  by the 
six-fold differential cross sections of the corresponding process,
using Monte Carlo   samples and a box method\cite{BOXMETHOD}.   

A maximum-likelihood fit reproduces the  $\rocroc$ content of Monte Carlo test samples within 
the statistical uncertainties. However, a large negative correlation exists between the 
$\roc\pi^\mp \pi^0$ and $\pipi \pizpiz $ (non-resonant) fitted fractions.
Both contributions are  necessary to fit the data. In the following,
only the $\rocroc$ content and the sum of the $\roc\pi^\mp \pi^0$ and $\pipi \pizpiz $ (non-resonant)
contributions are considered.

To check the quality of the fit, the $\pi^\pm\piz$ mass distributions of the data are compared 
with those of a mixture of Monte Carlo event samples from the processes (\ref{eq:eqn04}),  
in the proportion  determined by the fit.
The data and Monte Carlo distributions are in a good agreement over the entire $\q$ and $\mgg$ range,
an example is shown in  Figure~\ref{fig:composq2}.
Figure~\ref{fig:angles} shows a similar comparison for some angular variables.


\section {Results}

The  cross section, $\mathrm{\Delta \sigma_{ee}}$, of the process $\EE \to\EE\rocroc$
is measured as a function of  $\q$ and $\mgg$. The results are
listed in Tables~\ref{tbl:xsectq2}--\ref{tbl:xsectwgg_lep2}, together with the
efficiencies and the background fractions.
The statistical uncertainties, listed in the Tables, are those of the fit.
The differential cross section $d \sigma_{\mathrm{ee}} / d \q$, derived from 
$\mathrm{\Delta \sigma_{ee}}$, is listed in Table~\ref{tbl:xsectq2}.
When evaluating the differential cross section,
a correction  based on the $\q$-dependence of the $\rocroc$ Monte Carlo sample is applied, 
so as to assign the cross section value to the centre of the corresponding
$\q$-bin~\cite{BIN}.
We also give in Table~\ref{tbl:xsectq2} the sum of the differential cross sections of 
the sub-processes leading to $\roc\pi^\mp \pi^0$ and $\pipi \pizpiz $ (non-resonant) 
final states.

To evaluate the cross section  $\sigma_{\gamma\gamma}$ of the process
$\gamgam \to \rocroc$, the integral of the transverse photon luminosity function, 
$L_{TT}$, 
is computed for each $\q$ and $\mgg$  bin using the   program GALUGA \cite{GALUGA}, 
which performs $ O (\alpha ^4)$ QED calculations. The cross section $\sigma_{\gamma\gamma}$
is derived from  the measured cross section 
$\mathrm{\Delta \sigma_{ee}}$
using the relation
$\mathrm{\Delta \sigma_{ee}} = L_{TT}  \sigma_{\gamma\gamma}$. 
Thus,  $\sigma_{\gamma\gamma}$ represents
an effective cross section containing contributions from both transverse  and
longitudinal  photon polarisations.
The cross sections of the process $ \gamgam \to \rocroc$ are listed in  Table \ref{tbl:xsectq2} 
as a function of $\q$ and in Tables~\ref{tbl:xsectwgg_lep1} and \ref{tbl:xsectwgg_lep2}
as a function of $\mgg$.
The  sum of the cross sections of the processes $\gamgam \to \roc\pi^\mp \pi^0$ and 
$\gamgam \to \pipi \pizpiz $ (non-resonant) are also given in 
Tables~\ref{tbl:xsectwgg_lep1} and \ref{tbl:xsectwgg_lep2}.

To estimate the  systematic uncertainties on the measured cross sections,
we varied the selection of tracks and photons as well as the cuts in the event 
selection procedure, well beyond the resolution of the concerned variables.  
The contribution of the selection to the systematic uncertainties 
is in the range of $8\% - 18\%$.
The contribution of the fitting procedure is estimated by varying the
size and the occupancies of the boxes, as well as the binning of the data,  and is found to be in the range of
$10\% - 20\%$ for the fits in $\q$ bins and in the range of $10\% - 30\%$,
for the fits in bins of $\mgg$.
The systematic uncertainty of \mbox{$4\% - 8\%$}
 introduced by the background correction procedure is also included. 
Different form-factor parameterisations were used for re-weighting the Monte Carlo events
and the observed variations of the acceptance correspond to a systematic uncertainty 
in the range of \mbox{$2\% - 7\%$}.  

All  contributions are added in quadrature to obtain the 
systematic uncertainties quoted in Tables~\ref{tbl:xsectq2}--\ref{tbl:xsectwgg_lep2}.
The overall normalisation uncertainties related to the trigger efficiency determination
result in a $4\%$ relative uncertainty between the Z-pole and high-energy data.

\section {Discussion}

The cross section of the process $ \gamgam \to \rocroc$ as a function of $\mgg$
is plotted in Figure~\ref{fig:xsectwgg} together with the data from the L3 measurement
of  $\ro\ro$ production \cite{L3paper269}. Both cross sections have similar
dependence on  $\mgg$ and are of the same magnitude,
though the $\rocroc$ cross section is systematically higher than the  $\ro\ro$ one.
The ratio of the cross section for $\rocroc$ production relative to the $\ro\ro$ one, in
the kinematic region  
$ 1.1 \GeV \le  \mgg  \le 2.1 \GeV $ and $ 1.2 \GeV ^2 \le \q \le 8.5 \GeV^2$,
is
$  \sigma (\rocroc) / \sigma (\ro\ro)  = 1.81 \pm 0.47\,(\rm stat.)  \pm
0.22\,(\rm syst.)$,   compatible with the factor two expected for
an isospin \mbox{I = 0} state. 
These features of  the $\rho\rho$ production at high $\q$ are in contrast with 
the different $\mgg$ dependence and the observed suppression by about a factor five
of the $\rocroc$ production with respect to
 $\ro\ro$ in the data for $\q \approx 0$  and $\mgg < 2 \GeV$\cite{ARGUS,CELLO,TASSO}. 
A wide range of theoretical models was developed  \cite{ATTEMPTS} to explain 
this  feature,  but the reason of this behaviour is 
 still not   understood\cite{rosner}. The present measurement shows 
 that this peculiarity disappears at high $\q$.

The cross section of the process $ \gamgam \to \rocroc$ as a function of $\q$
is shown in Figure~\ref{fig:xsectq2}a, together with the L3 data  for $\roro$ production~\cite{L3paper269}.
 Both data sets have similar magnitude and  $\q$ dependence.
The $\rocroc$ production cross section is fitted with a form-factor parametrisation~\cite{GINZBURG}
  based on the generalised vector dominance model
(GVDM) \cite{GVDM}. This is found to reproduce well the $\q$-dependence of the data, with a value of
$\chi ^2 / \mathrm{d.o.f.} =  1.31 / 4 $.

Figure~\ref{fig:xsectq2}b shows the differential cross section $d \sigma_{\mathrm{ee}} / d \q$
of the reaction $\EE \to\EE\rocroc$,  together with the  L3 measurement for
$\EE \to\EE\roro$~\cite{L3paper269}. As  for $\roro$ production,
 the $\rocroc$ cross section is fitted to a form~\cite{DIEHL}
expected from QCD-based calculations\cite{DIEHLPAP}:
\begin{eqnarray}
\label{eq:eqn11}
 \mathrm{d} \sigma_{\mathrm{ee}} / \mathrm{d} \q \sim
\frac{1} { Q^n (\q + < \mgg >^2)^2} \, ,
\end{eqnarray}
with $< \mgg > = 1.91 \GeV $ being the average 
 $\mgg$-value in the $\q$ intervals  used.
The fit provides a good description of the $\q$ dependence of the data,
 with  $\chi ^2 / \mathrm{d.o.f.} =  0.31/3 $
 and 
an exponent $ n = 2.5 \pm 0.4$, to be compared with the expected value
$ n = 2$. Only statistical uncertainties are considered.
A common fit of the data   taken at the Z pole and at high energy
is justified by the almost constant values
of the photon polarisation parameter $\epsilon$, which determines
the energy dependence of the cross section. This result, together with
that of our previous fit to $\roro$ data,  $ n = 2.4 \pm 0.3$~\cite{L3paper269}, 
 provides further evidence 
for similar $\q$ dependence of the $\rocroc$  and $\roro$ production in the kinematic 
region (\ref{eq:rangelep1})--(\ref{eq:rangewgg}).

%
%
%

\newpage
\typeout{   }     
\typeout{Using author list for paper 287 -  }
\typeout{$Modified: Jul 15 2001 by smele $}
\typeout{!!!!  This should only be used with document option a4p!!!!}
\typeout{   }
%
%
%
%
%
%

\newcount\tutecount  \tutecount=0
\def\tutenum#1{\global\advance\tutecount by 1 \xdef#1{\the\tutecount}}
\def\tute#1{$^{#1}$}
\tutenum\aachen            
\tutenum\nikhef            
\tutenum\mich              
\tutenum\lapp              
\tutenum\basel             
\tutenum\lsu               
\tutenum\beijing           
\tutenum\bologna           
\tutenum\tata              
\tutenum\ne                
\tutenum\bucharest         
\tutenum\budapest          
\tutenum\mit               
\tutenum\panjab            
\tutenum\debrecen          
\tutenum\dublin            
\tutenum\florence          
\tutenum\cern              
\tutenum\wl                
\tutenum\geneva            
\tutenum\hamburg           
\tutenum\hefei             
\tutenum\lausanne          
\tutenum\lyon              
\tutenum\madrid            
\tutenum\florida           
\tutenum\milan             
\tutenum\moscow            
\tutenum\naples            
\tutenum\cyprus            
\tutenum\nymegen           
\tutenum\caltech           
\tutenum\perugia           
\tutenum\peters            
\tutenum\cmu               
\tutenum\potenza           
\tutenum\prince            
\tutenum\riverside         
\tutenum\rome              
\tutenum\salerno           
\tutenum\ucsd              
\tutenum\sofia             
\tutenum\korea             
\tutenum\taiwan            
\tutenum\tsinghua          
\tutenum\purdue            
\tutenum\psinst            
\tutenum\zeuthen           
\tutenum\eth               

{
\parskip=0pt
\noindent
{\bf The L3 Collaboration:}
\ifx\selectfont\undefined
 \baselineskip=10.8pt
 \baselineskip\baselinestretch\baselineskip
 \normalbaselineskip\baselineskip
 \ixpt
\else
 \fontsize{9}{10.8pt}\selectfont
\fi
\medskip
\tolerance=10000
\hbadness=5000
\raggedright
\hsize=162truemm\hoffset=0mm
\def\r{\rlap,}
\noindent

P.Achard\r\tute\geneva\ 
O.Adriani\r\tute{\florence}\ 
M.Aguilar-Benitez\r\tute\madrid\ 
J.Alcaraz\r\tute{\madrid}\ 
G.Alemanni\r\tute\lausanne\
J.Allaby\r\tute\cern\
A.Aloisio\r\tute\naples\ 
M.G.Alviggi\r\tute\naples\
H.Anderhub\r\tute\eth\ 
V.P.Andreev\r\tute{\lsu,\peters}\
% ...
\newpage
\begin{list}{A}{\itemsep=0pt plus 0pt minus 0pt\parsep=0pt plus 0pt minus 0pt
                \topsep=0pt plus 0pt minus 0pt}
\item[\aachen]
 III. Physikalisches Institut, RWTH, D-52056 Aachen, Germany$^{\S}$
\item[\nikhef] National Institute for High Energy Physics, NIKHEF, 
     and University of Amsterdam, NL-1009 DB Amsterdam, The Netherlands
\item[\mich] University of Michigan, Ann Arbor, MI 48109, USA
\item[\lapp] Laboratoire d'Annecy-le-Vieux de Physique des Particules, 
     LAPP,IN2P3-CNRS, BP 110, F-74941 Annecy-le-Vieux CEDEX, France
\item[\basel] Institute of Physics, University of Basel, CH-4056 Basel,
     Switzerland
\item[\lsu] Louisiana State University, Baton Rouge, LA 70803, USA
\item[\beijing] Institute of High Energy Physics, IHEP, 
  100039 Beijing, China$^{\triangle}$ 
\item[\bologna] University of Bologna and INFN-Sezione di Bologna, 
     I-40126 Bologna, Italy
\item[\tata] Tata Institute of Fundamental Research, Mumbai (Bombay) 400 005, India
\item[\ne] Northeastern University, Boston, MA 02115, USA
\item[\bucharest] Institute of Atomic Physics and University of Bucharest,
     R-76900 Bucharest, Romania
\item[\budapest] Central Research Institute for Physics of the 
     Hungarian Academy of Sciences, H-1525 Budapest 114, Hungary$^{\ddag}$
\item[\mit] Massachusetts Institute of Technology, Cambridge, MA 02139, USA
\item[\panjab] Panjab University, Chandigarh 160 014, India
\item[\debrecen] KLTE-ATOMKI, H-4010 Debrecen, Hungary$^\P$
\item[\dublin] Department of Experimental Physics,
  University College Dublin, Belfield, Dublin 4, Ireland
\item[\florence] INFN Sezione di Firenze and University of Florence, 
     I-50125 Florence, Italy
\item[\cern] European Laboratory for Particle Physics, CERN, 
     CH-1211 Geneva 23, Switzerland
\item[\wl] World Laboratory, FBLJA  Project, CH-1211 Geneva 23, Switzerland
\item[\geneva] University of Geneva, CH-1211 Geneva 4, Switzerland
\item[\hamburg] University of Hamburg, D-22761 Hamburg, Germany
\item[\hefei] Chinese University of Science and Technology, USTC,
      Hefei, Anhui 230 029, China$^{\triangle}$
\item[\lausanne] University of Lausanne, CH-1015 Lausanne, Switzerland
\item[\lyon] Institut de Physique Nucl\'eaire de Lyon, 
     IN2P3-CNRS,Universit\'e Claude Bernard, 
     F-69622 Villeurbanne, France
\item[\madrid] Centro de Investigaciones Energ{\'e}ticas, 
     Medioambientales y Tecnol\'ogicas, CIEMAT, E-28040 Madrid,
     Spain${\flat}$ 
\item[\florida] Florida Institute of Technology, Melbourne, FL 32901, USA
\item[\milan] INFN-Sezione di Milano, I-20133 Milan, Italy
\item[\moscow] Institute of Theoretical and Experimental Physics, ITEP, 
     Moscow, Russia
\item[\naples] INFN-Sezione di Napoli and University of Naples, 
     I-80125 Naples, Italy
\item[\cyprus] Department of Physics, University of Cyprus,
     Nicosia, Cyprus
\item[\nymegen] University of Nijmegen and NIKHEF, 
     NL-6525 ED Nijmegen, The Netherlands
\item[\caltech] California Institute of Technology, Pasadena, CA 91125, USA
\item[\perugia] INFN-Sezione di Perugia and Universit\`a Degli 
     Studi di Perugia, I-06100 Perugia, Italy   
\item[\peters] Nuclear Physics Institute, St. Petersburg, Russia
\item[\cmu] Carnegie Mellon University, Pittsburgh, PA 15213, USA
\item[\potenza] INFN-Sezione di Napoli and University of Potenza, 
     I-85100 Potenza, Italy
\item[\prince] Princeton University, Princeton, NJ 08544, USA
\item[\riverside] University of Californa, Riverside, CA 92521, USA
\item[\rome] INFN-Sezione di Roma and University of Rome, ``La Sapienza",
     I-00185 Rome, Italy
\item[\salerno] University and INFN, Salerno, I-84100 Salerno, Italy
\item[\ucsd] University of California, San Diego, CA 92093, USA
\item[\sofia] Bulgarian Academy of Sciences, Central Lab.~of 
     Mechatronics and Instrumentation, BU-1113 Sofia, Bulgaria
\item[\korea]  The Center for High Energy Physics, 
     Kyungpook National University, 702-701 Taegu, Republic of Korea
\item[\taiwan] National Central University, Chung-Li, Taiwan, China
\item[\tsinghua] Department of Physics, National Tsing Hua University,
      Taiwan, China
\item[\purdue] Purdue University, West Lafayette, IN 47907, USA
\item[\psinst] Paul Scherrer Institut, PSI, CH-5232 Villigen, Switzerland
\item[\zeuthen] DESY, D-15738 Zeuthen, Germany
\item[\eth] Eidgen\"ossische Technische Hochschule, ETH Z\"urich,
     CH-8093 Z\"urich, Switzerland
\item[\S]  Supported by the German Bundesministerium 
        f\"ur Bildung, Wissenschaft, Forschung und Technologie.
\item[\ddag] Supported by the Hungarian OTKA fund under contract
numbers T019181, F023259 and T037350.
\item[\P] Also supported by the Hungarian OTKA fund under contract
  number T026178.
\item[$\flat$] Supported also by the Comisi\'on Interministerial de Ciencia y 
        Tecnolog{\'\i}a.
\item[$\sharp$] Also supported by CONICET and Universidad Nacional de La Plata,
        CC 67, 1900 La Plata, Argentina.
\item[$\triangle$] Supported by the National Natural Science
  Foundation of China.
\end{list}
}
\vfill


\newpage

\begin{table}[htb]
\begin{center}
\begin{sideways}
\begin{minipage}[b]{\textheight}
\begin{center}
\begin{tabular}{|c|c|c|l|l|c|l|}
\hline
$ \q$-range &  $\varepsilon$ & $\mathit{Bg}$ & $~~~~~~\Delta \sigma_{ee}$ [ pb ] & $ \;\, d\,\sigma_{ee}/d\,\q$ [ pb\,/$\GeV^2 \,]$ & 
$ \sigma_{\gamma\gamma}$ [ nb ]  & $ \, d\,\sigma_{ee}/d\,\q$ [ pb\,/$\GeV^2 \,]$ \\
 $[\;\GeV^2 \;]$ & [ \% ] & [ \% ] & $~~~~~~~~~~~~\rocroc$ & $~~~~~~~~~~~~~~\rocroc$ &  $\rocroc$ &  $ ~~\roc\pi^\mp \pi^0 + \pipi\pi^0\pi^0$  \\ \hline

\phantom{0}1.2 -- \phantom{0}2.2  & 3.7  & 19         & $   6.30            \pm 1.63\phantom{0}  \pm 1.07\phantom{0} $ & $ 6.06\phantom{0} \pm 1.57\phantom{00} \pm 1.03\phantom{00} $  & $ 5.12 \pm 1.32 \pm 0.87 $ & $ 8.63\phantom{0} \pm 1.71\phantom{00} \pm 1.21\phantom{00} $ \\ \hline
\phantom{0}2.2 -- \phantom{0}3.5  & 5.0  & 18         & $   2.57            \pm 0.96\phantom{0}  \pm 0.58\phantom{0} $ & $ 1.85\phantom{0} \pm 0.69\phantom{00} \pm 0.41\phantom{00} $  & $ 3.33 \pm 1.24 \pm 0.75 $ & $ 3.51\phantom{0} \pm 0.80\phantom{00} \pm 0.54\phantom{00} $ \\ \hline
\phantom{0}3.5 -- \phantom{0}8.5  & 5.6  & 18         & $   2.11            \pm 0.81\phantom{0}  \pm 0.41\phantom{0} $ & $ 0.31\phantom{0} \pm 0.12\phantom{00} \pm 0.06\phantom{00} $  & $ 1.98 \pm 0.77 \pm 0.38 $ & $ 0.53\phantom{0} \pm 0.13\phantom{00} \pm 0.07\phantom{00} $ \\ \hline
\phantom{0}8.8 -- 14.0            & 5.6  & 16         & $   0.38            \pm 0.135            \pm 0.072 $           & $ 0.067            \pm 0.024\phantom{0} \pm 0.013\phantom{0}  $  & $ 0.74 \pm 0.26 \pm 0.14 $ & $ 0.14\phantom{0} \pm 0.029\phantom{0} \pm 0.019\phantom{0} $ \\ \hline
14.0 -- 30.0                      & 6.3  & 17         & $   0.23            \pm 0.099            \pm 0.060 $           & $ 0.011            \pm 0.0046           \pm 0.0029           $  & $ 0.40 \pm 0.17 \pm 0.10 $ & $ 0.024           \pm 0.0058           \pm 0.0041           $ \\ \hline
  
\end{tabular}
\end{center}

\caption{Detection efficiencies, $\varepsilon$,  background fractions, $\mathit{Bg}$,  
         and measured production cross sections as a function of  $\q$ for $1.1 \GeV < \mgg < 3 \GeV$
       	 for   Z-pole and high-energy data.
         The first uncertainties are statistical, the second systematic.
         The values of the differential cross section are  corrected to the centre of each bin.
         }
\label{tbl:xsectq2}
\end{minipage}
\end{sideways}
\end{center}
\end{table}

\begin{table*}[ht]
\begin{center}
\begin{tabular}{|c|c|c|c|c|c|}
\hline
$  \mgg$-range & $\varepsilon$&  $\mathit{Bg}$ &  $ \Delta \sigma_{ee}$ [ pb ] & 
$ \sigma_{\gamma\gamma}$ [ nb ]  & $ \sigma_{\gamma\gamma}$ [ nb ] \\
$[\;\GeV \;\;]$ & [ \% ] & [ \% ] &  $\rocroc$ &  $\rocroc$ &  $ \roc\pi^\mp \pi^0 + \pipi\pi^0\pi^0$  \\ \hline
1.1 -- 1.5    & 3.2 & 28 & $ 3.09 \pm 1.18  \pm 0.96  $  & $ 3.99 \pm 1.53 \pm 1.24 $ & $ 7.51 \pm 1.78 \pm 1.43 $ \\ \hline
1.5 -- 1.8    & 4.2 & 17 & $ 3.67 \pm 1.04  \pm 0.55  $  & $ 6.84 \pm 1.93 \pm 1.03 $ & $ 7.90 \pm 2.03 \pm 1.15 $ \\ \hline
1.8 -- 2.1    & 4.6 & 14 & $ 2.79 \pm 0.81  \pm 0.39  $  & $ 5.62 \pm 1.63 \pm 0.79 $ & $ 6.57 \pm 1.74 \pm 0.88 $ \\ \hline
2.1 -- 3.0    & 5.3 & 14 & $ 1.95 \pm 0.69  \pm 0.38  $  & $ 1.55 \pm 0.55 \pm 0.30 $ & $ 3.87 \pm 0.74 \pm 0.50 $ \\ \hline
\end{tabular}
\caption{Detection efficiencies, $\varepsilon$,  background fractions, $\mathit{Bg}$,
         and measured production cross sections 
         as a function of  $\mgg$, for
         $1.2 \GeV^2 < \q < 8.5 \GeV^2$, for  the Z-pole data, 
         together with  the cross sections of the reactions
         $ \EE \to \EE \rocroc$, $\gamgam \to \rocroc$ 
        	and the sum of the cross sections of the processes 
         $\gamgam \to \roc\pi^\mp \pi^0 $ and $\gamgam \to \pipi\pi^0\pi^0$(non-resonant). 
         The first uncertainties are statistical, the second systematic.
         }
\label{tbl:xsectwgg_lep1}
\end{center}
\end{table*}


\begin{table*}[ht]
\begin{center}
\begin{tabular}{|c|c|c|c|c|c|c|}
\hline
$ \mgg$-range & $\varepsilon$ &  $\mathit{Bg}$ &  $ \Delta \sigma_{ee}$ [ pb ] & 
$   \sigma_{\gamma\gamma}$ [ nb ]  & $\sigma_{\gamma\gamma}$ [ nb ] \\
$[\;\GeV \;\;]$& [ \% ] & [ \% ] & $\rocroc$  & $\rocroc$ & $ \roc\pi^\mp \pi^0 + \pipi\pi^0\pi^0$ \\ \hline
1.1 -- 1.7    & 4.9  & 24 & $ 0.218 \pm 0.109 \pm 0.059 $   & $ 0.62 \pm 0.31 \pm 0.17 $ & $ 1.12 \pm 0.37 \pm 0.25 $ \\ \hline
1.7 -- 2.2    & 6.1  & 16 & $ 0.272 \pm 0.119 \pm 0.082 $   & $ 0.95 \pm 0.42 \pm 0.29 $ & $ 1.52 \pm 0.48 \pm 0.33 $ \\ \hline
2.2 -- 3.0    & 6.4  & 11 & $ 0.121 \pm 0.078 \pm 0.040 $   & $ 0.27 \pm 0.18 \pm 0.09 $ & $ 1.10 \pm 0.26 \pm 0.21 $ \\ \hline
\end{tabular}
\caption{Detection efficiency, $\varepsilon$,  background fractions, $\mathit{Bg}$,
         and measured production cross sections
         as a function of  $\mgg$, for
         $8.8 \GeV^2 < \q < 30  \GeV^2$, for the high energy data.
         together with the cross sections of the reactions
         $ \EE \to \EE \rocroc$, $\gamgam \to \rocroc$ 
        	and the sum of the cross sections of the processes 
         $\gamgam \to \roc\pi^\mp \pi^0 $ and $\gamgam \to \pipi\pi^0\pi^0$(non-resonant). 
         The first uncertainties are statistical, the second systematic.
         }
\label{tbl:xsectwgg_lep2}
\end{center}
\end{table*}

%
%
%
\clearpage
\vfil

\clearpage
\vfil


\clearpage


\clearpage
\vskip +1.5cm

\hskip -0.5cm
\begin{minipage}{.5cm}
\begin{center}
\large
\rotatebox{90}{Entries / 80 MeV}\\
\normalsize
\end{center}
\end{minipage}\begin{minipage}{16.5cm}

{\epsfig{file=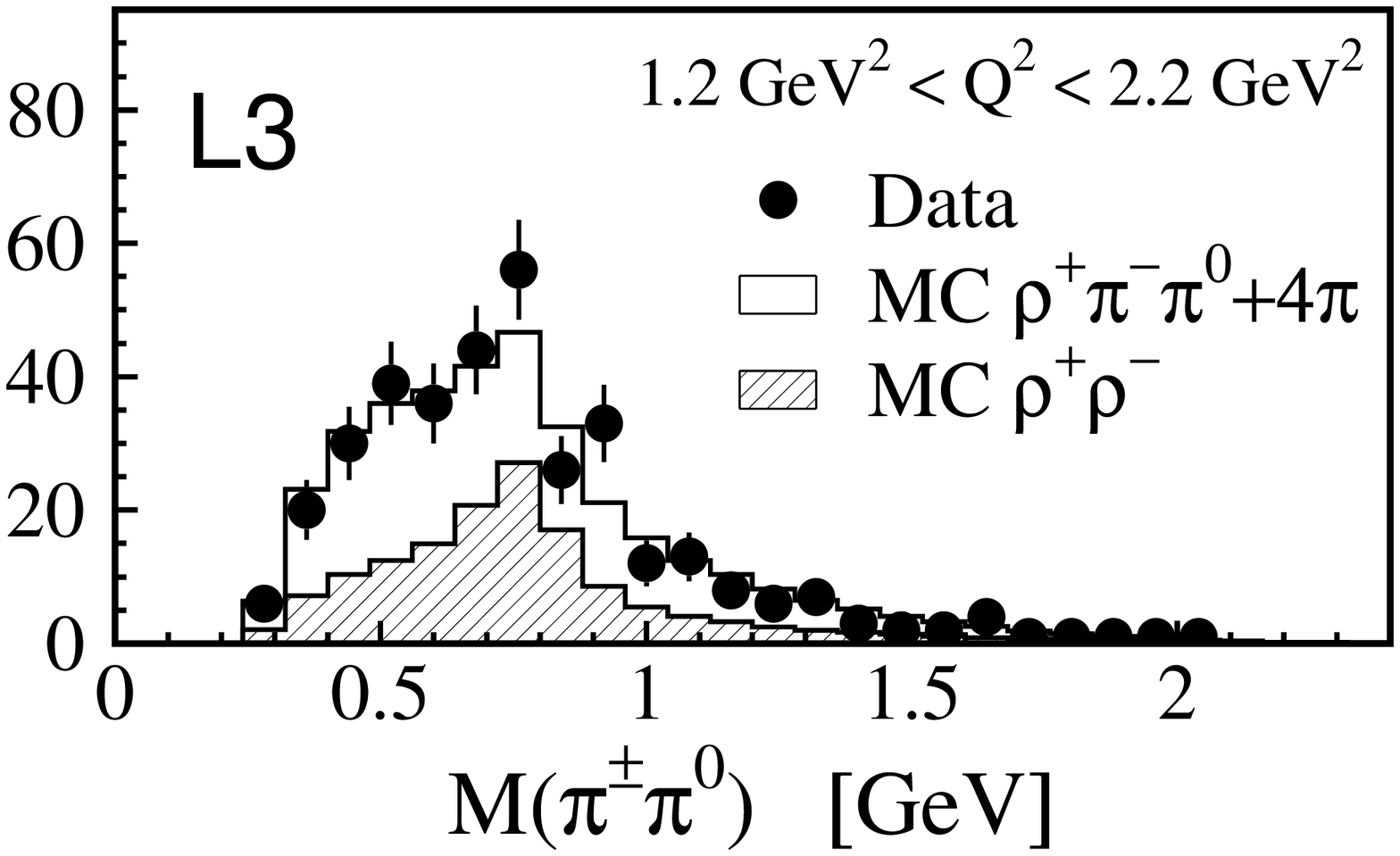,width=0.49\linewidth}}
{\epsfig{file=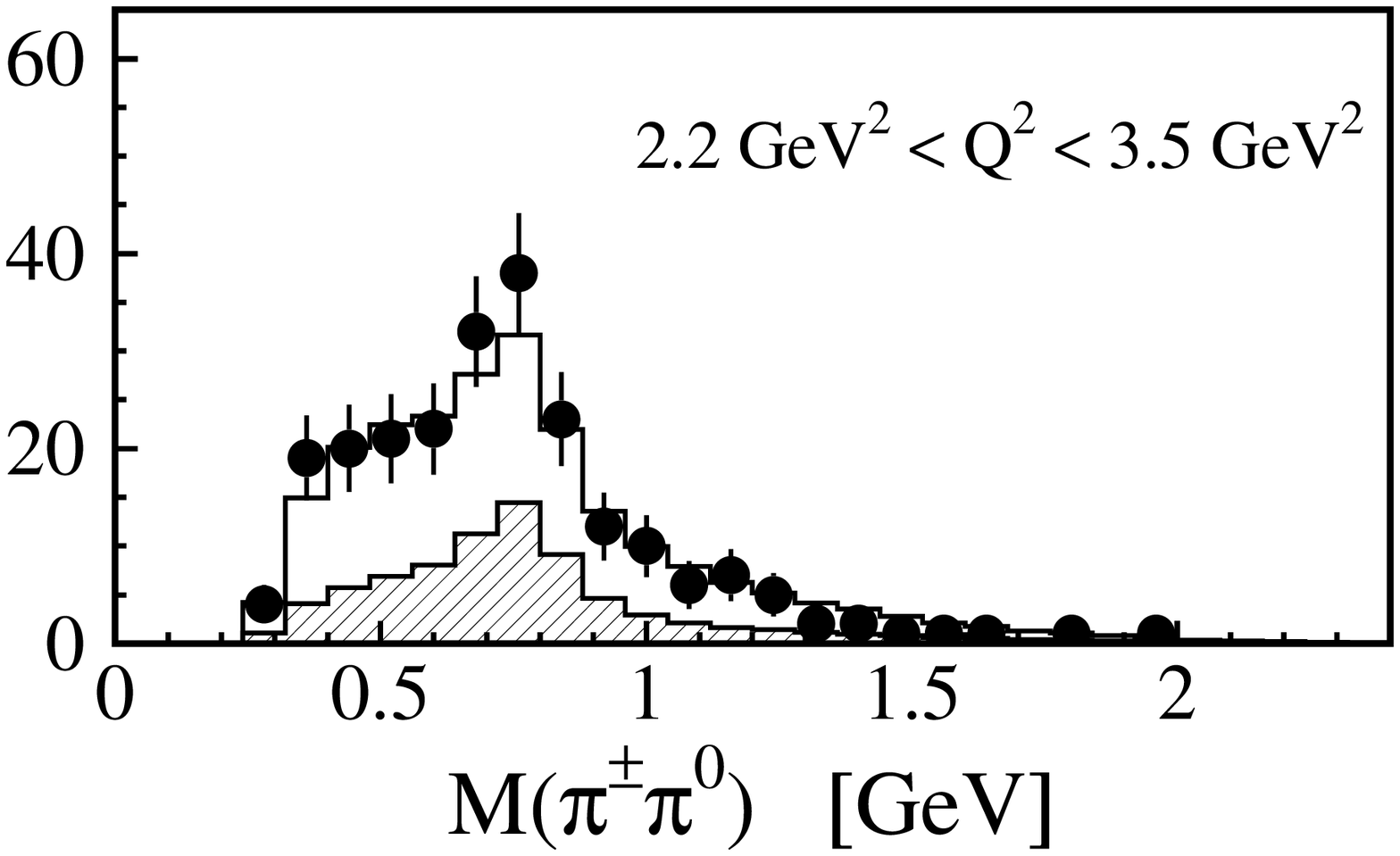,width=0.49\linewidth}}
{\epsfig{file=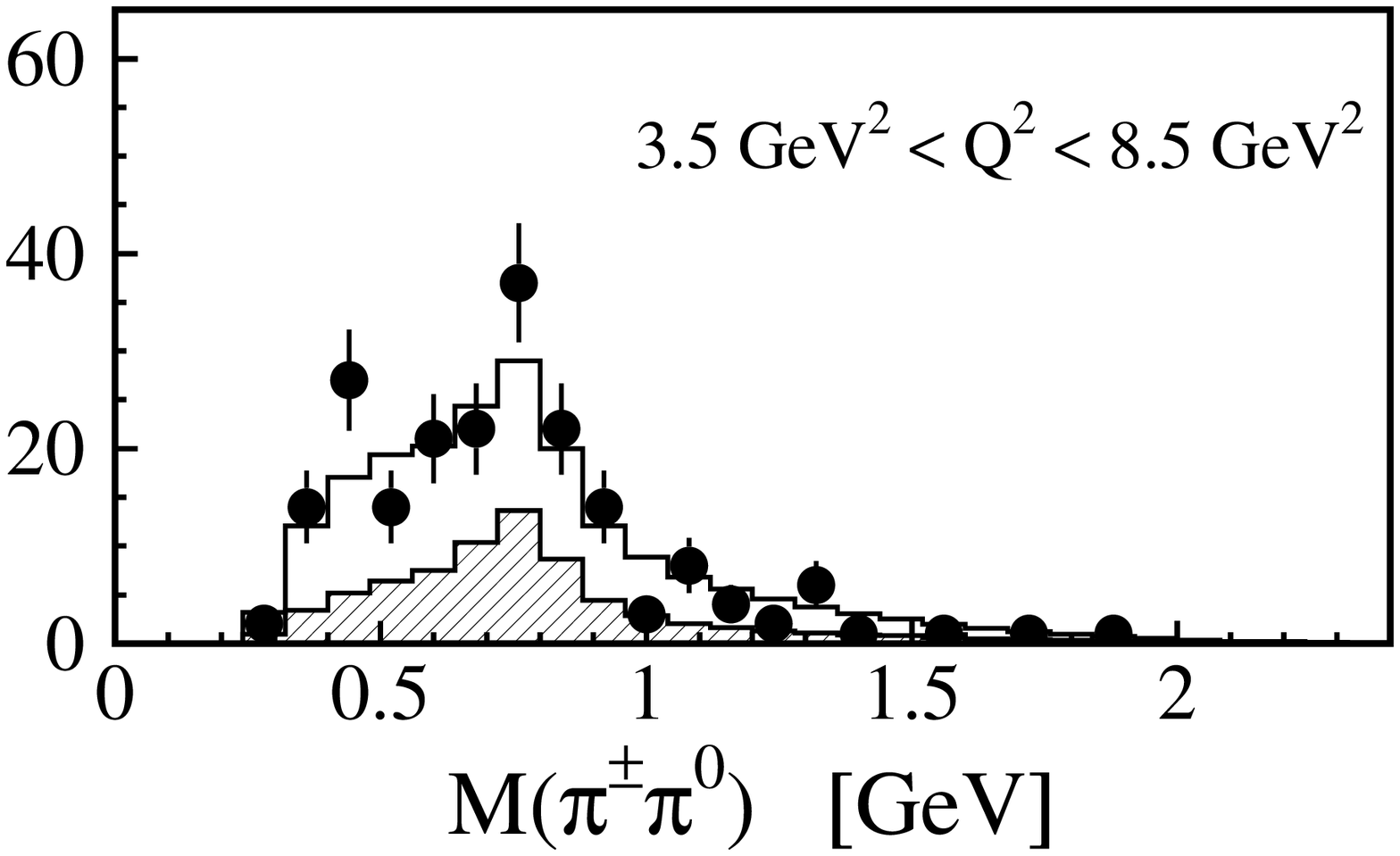,width=0.49\linewidth}}
{\epsfig{file=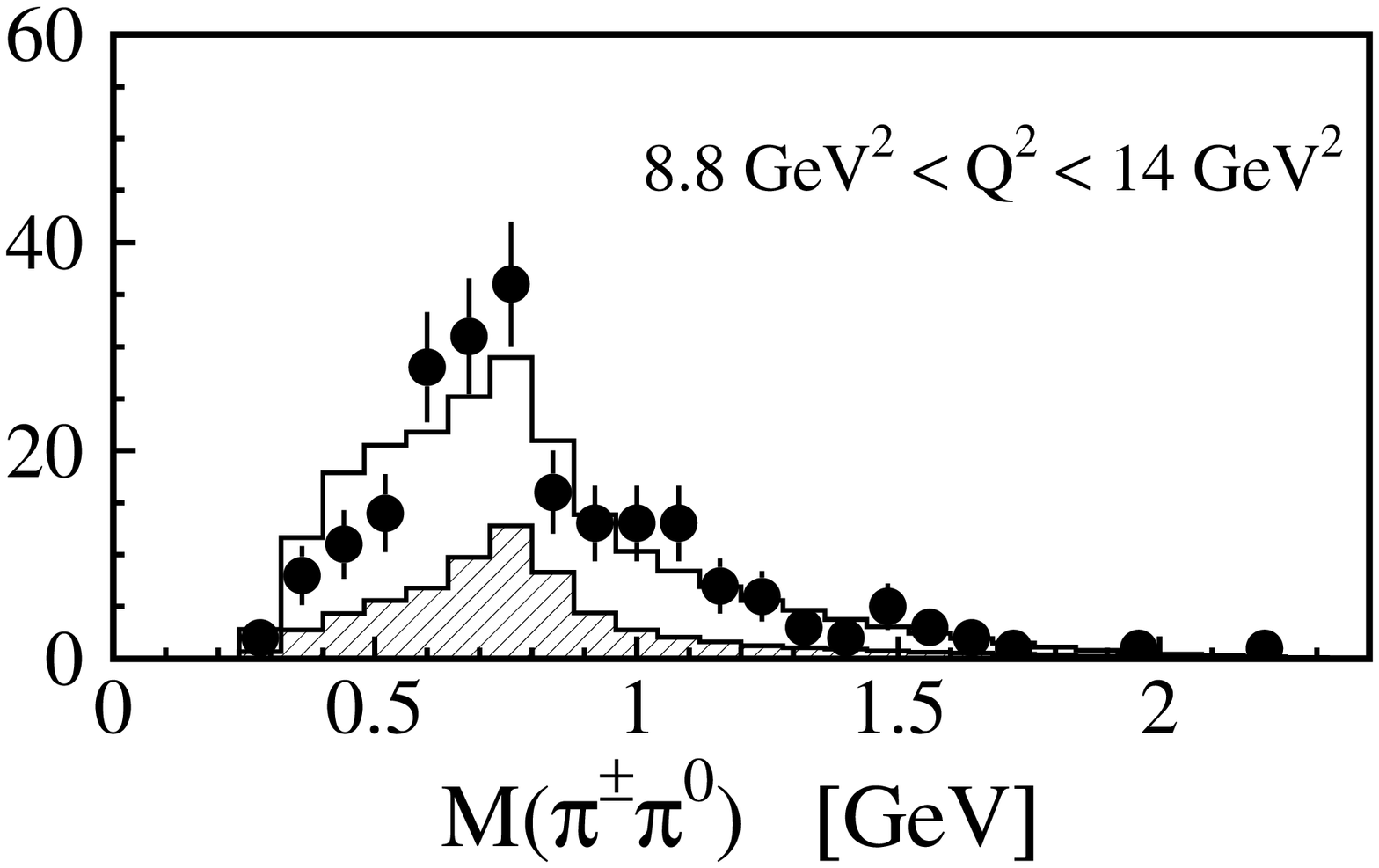,width=0.49\linewidth}}
{\epsfig{file=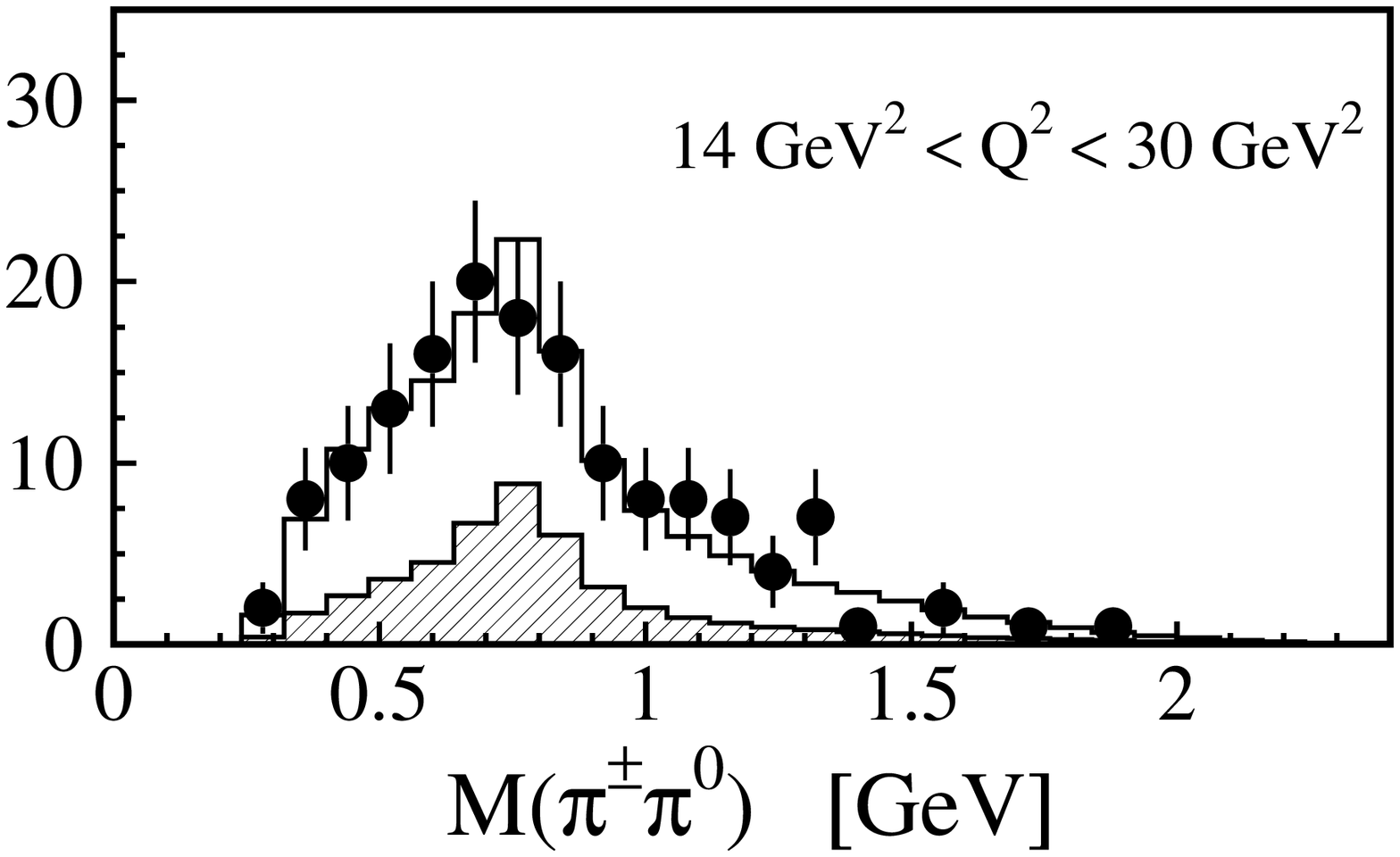,width=0.49\linewidth}}
{\epsfig{file=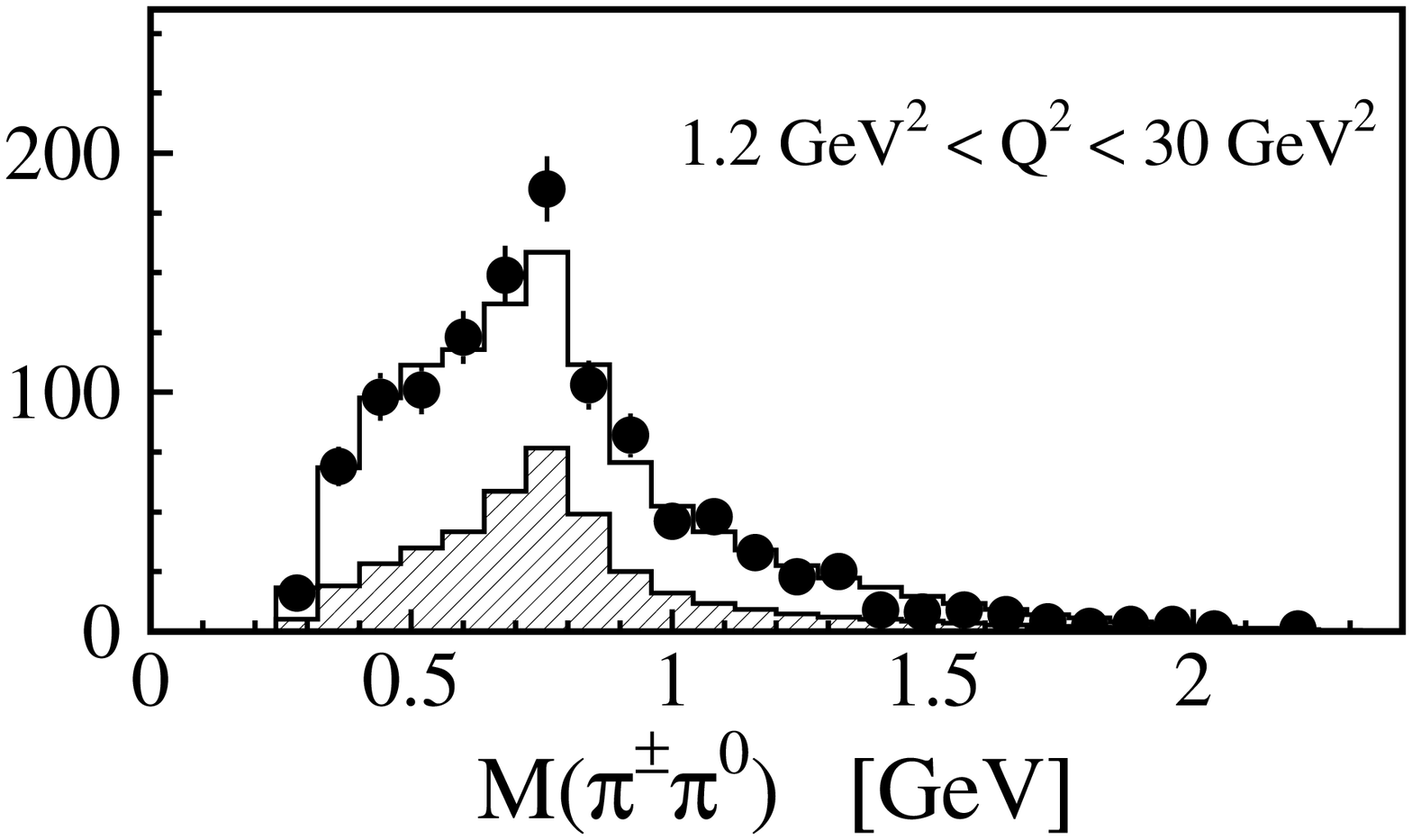,width=0.49\linewidth}}
\end{minipage}

\begin{figure}[ht]
\caption[]{Effective mass distributions of  $\pi^\pm \piz$ combinations (four entries per event)
           for events  with $1.1 \GeV < \mgg < 3 \GeV$
           in the fitted $\q$ intervals.
           The points represent the data, the hatched area shows  the $\rocroc$ component
           and the open area shows the sum of
           $\roc\pi^\mp \pi^0$ and $\pipi\pi^0\pi^0$ (non-resonant) components.
           The fraction of the different components are determined by the fit and the
           total normalisation is to the number of the events.
           The plot for the entire  $\q$ range, $1.2 \GeV^2 < \q < 30 \GeV^2$, is the sum of 
           the distributions of the five fitted $\q$ intervals. 
           }
\label{fig:composq2}
\end{figure}
\vfil

\vfil


  \begin{figure} [p]
  \begin{center}
    \mbox{\epsfig{file=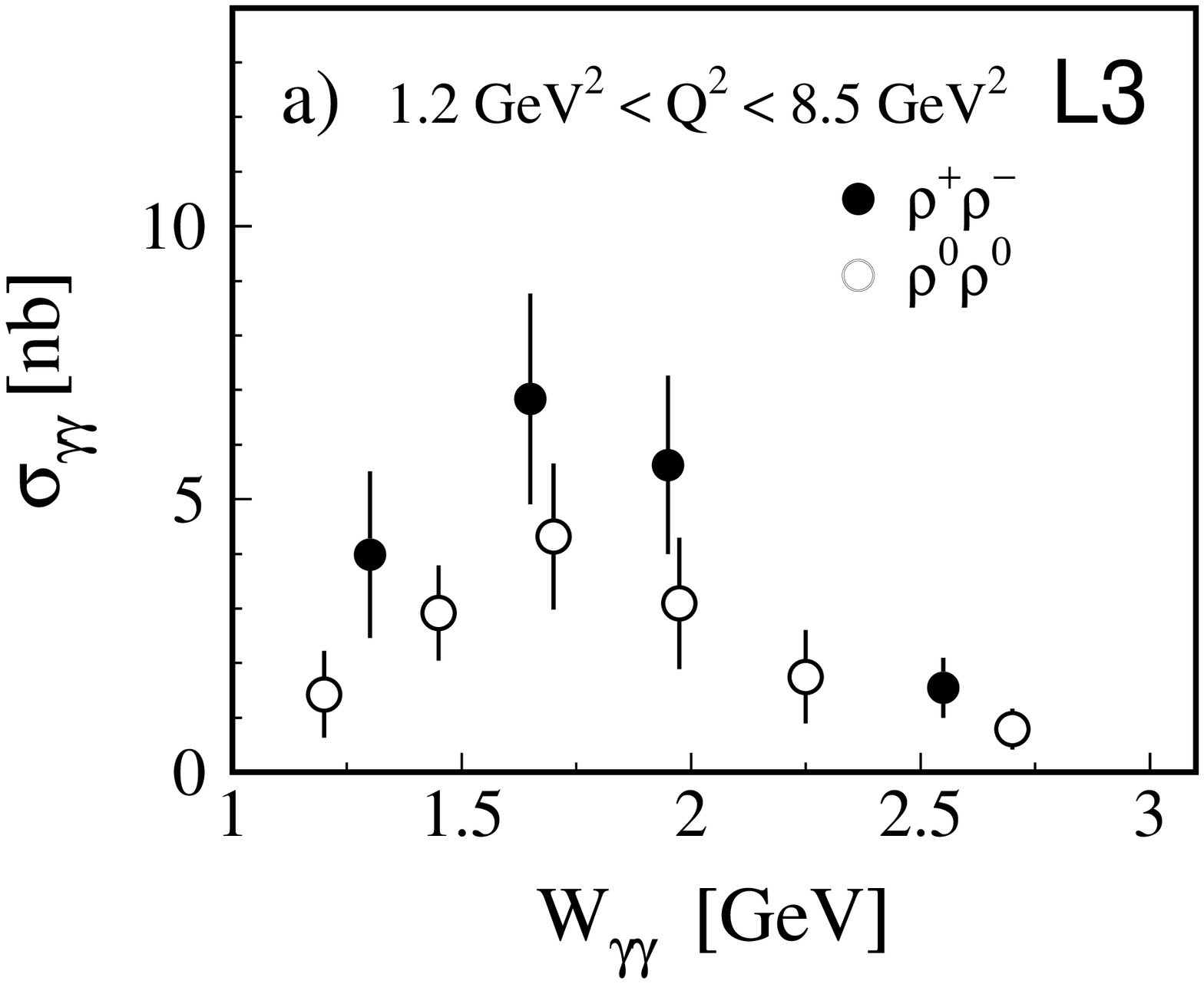,width=0.49\textwidth}}
    \mbox{\epsfig{file=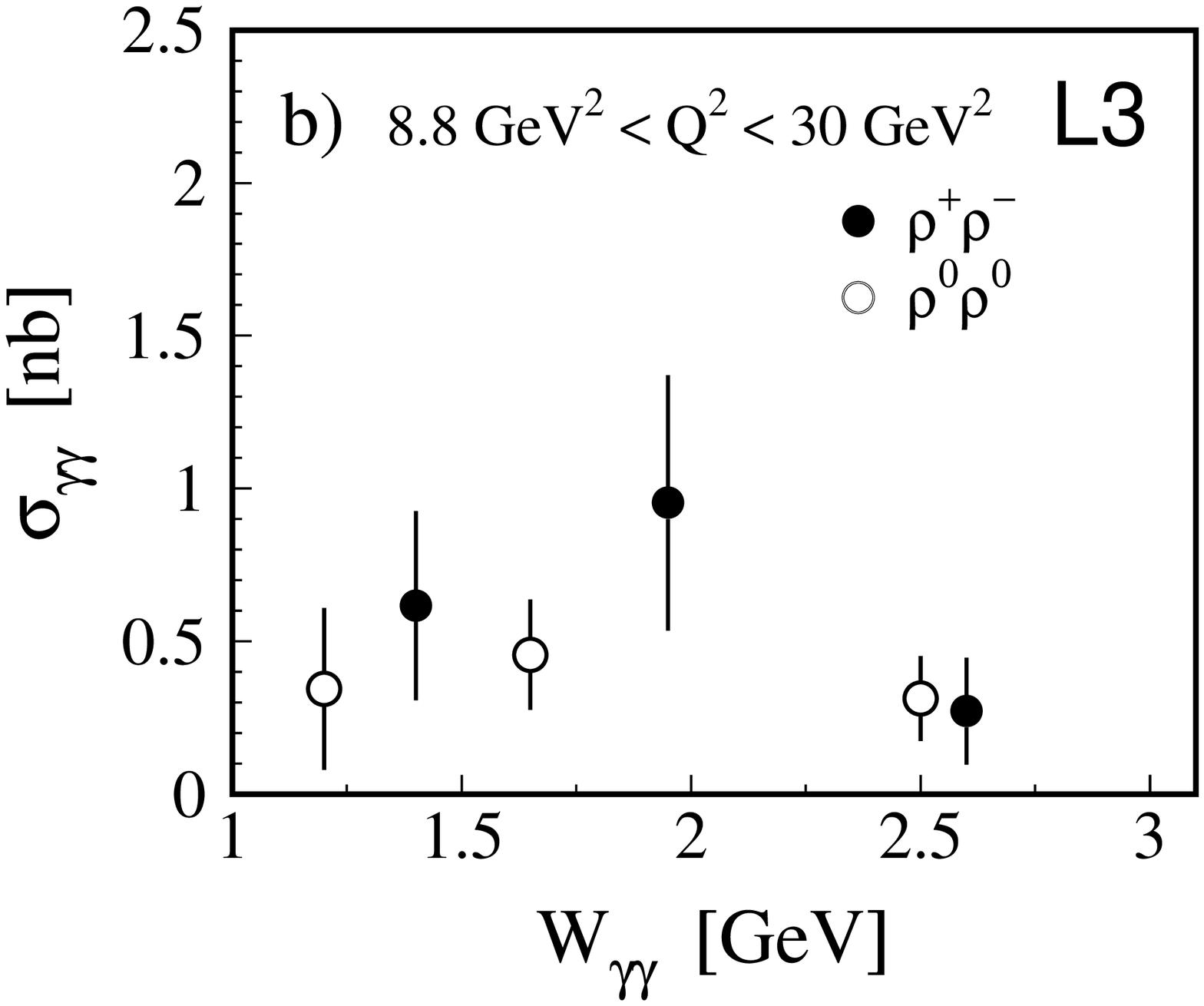,width=0.49\textwidth}}
  \end{center} 
  \caption[]{
            Cross section of the process $\gamgam\to\rho\rho$
            as  function of  $\mgg$,  for
            a) $1.2 \GeV^2 < \q < 8.5 \GeV^2$ and
            b) $8.8 \GeV^2 < \q < 30 \GeV^2$.
         The full points show the results from this measurement,
         the open points show the results from the L3 measurement
         of $\ro\ro$ production \protect\cite{L3paper269},
         the  bars show the statistical uncertainties.
           }
\label{fig:xsectwgg}
\end{figure}
%
%

  \begin{figure} [p]
  \begin{center}
    \mbox{\epsfig{file=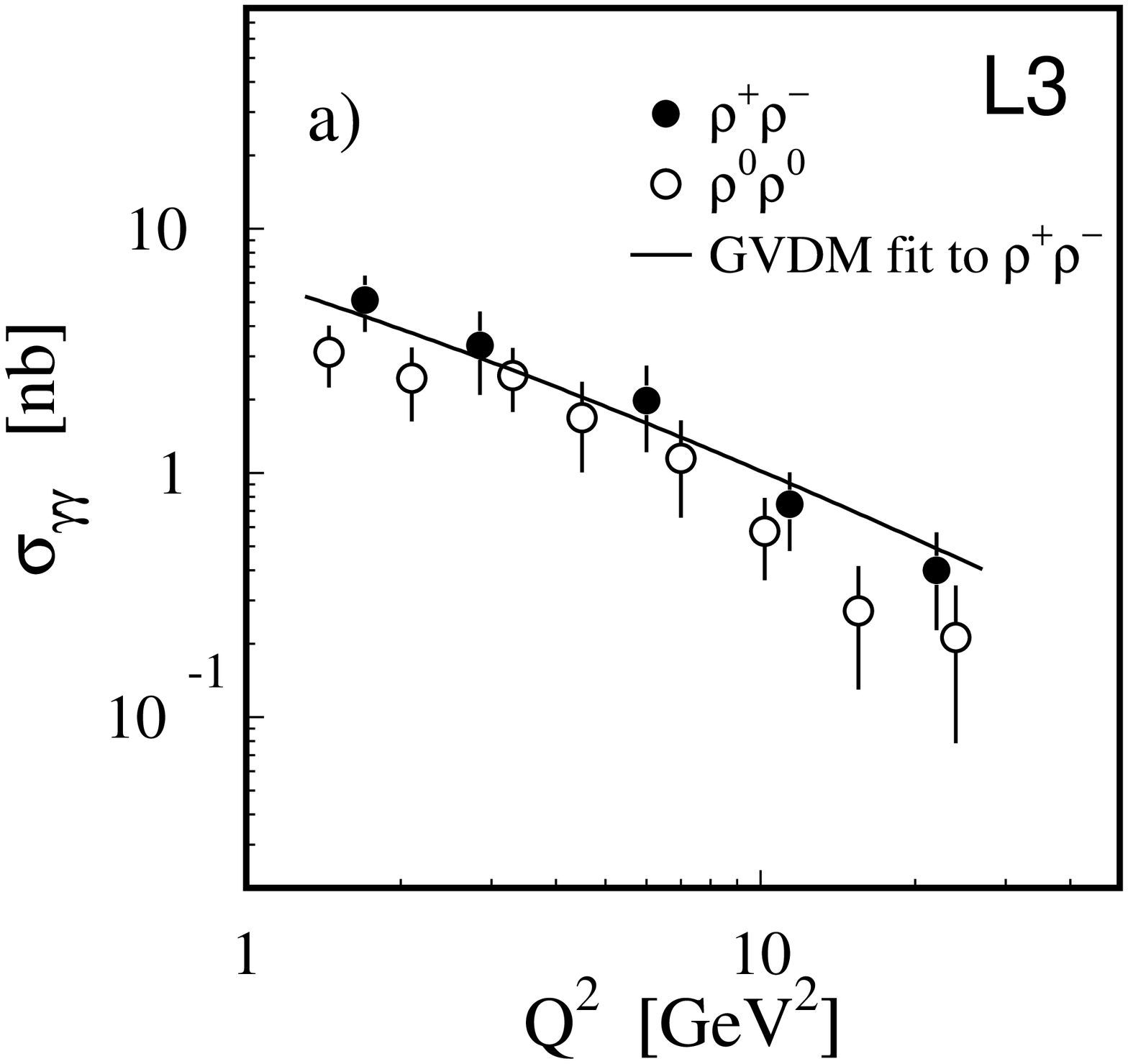,width=0.49\textwidth}}
    \mbox{\epsfig{file=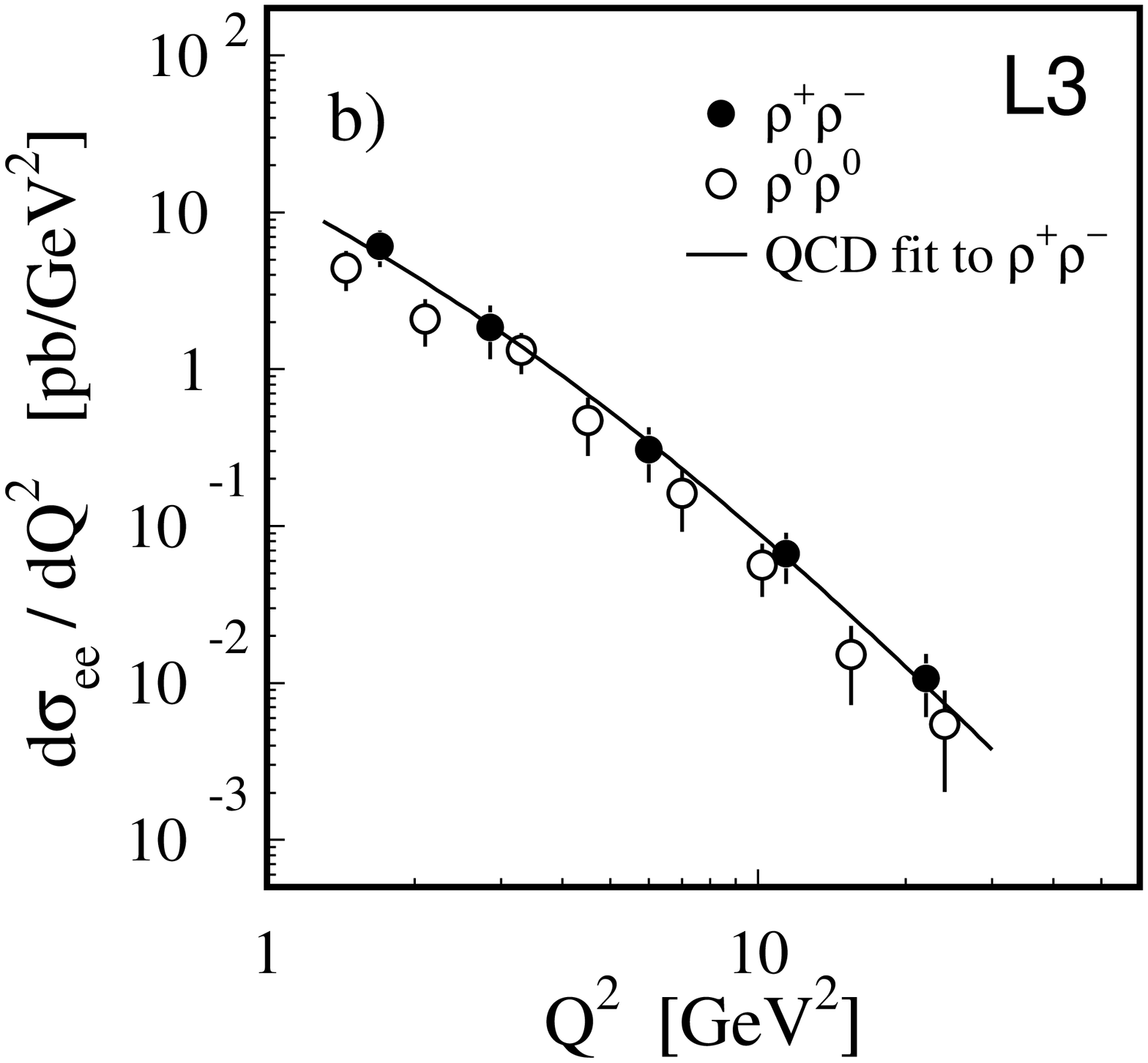,width=0.49\textwidth}}
  \vspace{-3mm}
  \end{center}
 \caption{The $\rho\rho$ production cross section as a function of
         $\q$, for $1.1 \GeV < \mgg < 3 \GeV$:
         a) cross section of the process $\gamgam \to \rho\rho$ and
         b) differential cross section of the process
         $ \EE \to \EE \rho\rho$.
         The full points show the results from this measurement,
         the open points show the results from the L3 measurement
         of $\ro\ro$ production \protect\cite{L3paper269},
         the  bars show the statistical uncertainties.
         The line in a) represents the result of a fit based on
         the generalised vector-meson dominance model
         \protect\cite{GINZBURG}.
         The line in b) represents the result of a fit to a form
         expected from QCD calculations.
           }
\label{fig:xsectq2}
\end{figure}
\vfil

%
%
%
%

\end{document}